\documentclass[twocolumn,amsmath,amssymb,floatfix,superscriptaddress]{revtex4-2}
\usepackage{bm,graphicx,url,epsf,color}
\usepackage{wasysym}
\usepackage{MnSymbol}
\usepackage[normalem]{ulem}

\begin{document}

\title{Dynamical Coulomb blockade under a temperature bias}

\author{H. Duprez}
\affiliation{Universit\'e Paris-Saclay, CNRS, Centre de Nanosciences et de Nanotechnologies (C2N), 91120 Palaiseau, France}
\author{F. Pierre}
\email[e-mail: ]{frederic.pierre@c2n.upsaclay.fr}
\affiliation{Universit\'e Paris-Saclay, CNRS, Centre de Nanosciences et de Nanotechnologies (C2N), 91120 Palaiseau, France}
\author{E. Sivre}
\affiliation{Universit\'e Paris-Saclay, CNRS, Centre de Nanosciences et de Nanotechnologies (C2N), 91120 Palaiseau, France}
\author{A. Aassime}
\affiliation{Universit\'e Paris-Saclay, CNRS, Centre de Nanosciences et de Nanotechnologies (C2N), 91120 Palaiseau, France}
\author{F.D. Parmentier}
\affiliation{Universit\'e Paris-Saclay, CNRS, Centre de Nanosciences et de Nanotechnologies (C2N), 91120 Palaiseau, France}
\author{A. Cavanna}
\affiliation{Universit\'e Paris-Saclay, CNRS, Centre de Nanosciences et de Nanotechnologies (C2N), 91120 Palaiseau, France}
\author{A. Ouerghi}
\affiliation{Universit\'e Paris-Saclay, CNRS, Centre de Nanosciences et de Nanotechnologies (C2N), 91120 Palaiseau, France}
\author{U. Gennser}
\affiliation{Universit\'e Paris-Saclay, CNRS, Centre de Nanosciences et de Nanotechnologies (C2N), 91120 Palaiseau, France}
\author{I. Safi}
\affiliation{Universit\'e Paris-Saclay, CNRS, Laboratoire de Physique des Solides, 91405 Orsay, France}
\author{C. Mora}
\affiliation{Universit\'e  de  Paris, Laboratoire  Mat\'eriaux  et  Ph\'enom\`enes  Quantiques, CNRS, 75013  Paris,  France}
\affiliation{Dahlem Center for Complex Quantum Systems and Fachbereich Physik, Freie Universit\"at Berlin, 14195, Berlin, Germany}
\author{A. Anthore}
\affiliation{Universit\'e Paris-Saclay, CNRS, Centre de Nanosciences et de Nanotechnologies (C2N), 91120 Palaiseau, France}
\affiliation{Universit\'{e} de Paris, C2N, 91120 Palaiseau, France}

\begin{abstract}
We observe and comprehend the dynamical Coulomb blockade suppression of the electrical conductance across an electronic quantum channel submitted to a temperature difference.
A broadly tunable, spin-polarized Ga(Al)As quantum channel is connected on-chip, through a micron-scale metallic node, to a linear $RC$ circuit.
The latter is made up of the node's geometrical capacitance $C$ in parallel with an adjustable resistance $R\in \{1/2,1/3,1/4\}\times h/e^2$ formed by 2--4 quantum Hall channels.
The system is characterized by three temperatures: a temperature of the electrons in the large electrodes ($T$) and in the node ($T_\mathrm{node}$), and a temperature of the electromagnetic modes of the $RC$ circuit ($T_\mathrm{env}$).
The temperature in the node is selectively increased by local Joule dissipation, and characterized from current fluctuations.
For a quantum channel in the tunnel regime, a close match is found between conductance measurements and tunnel dynamical Coulomb blockade theory.
In the opposite near ballistic regime, we develop a theory that accounts for different electronic and electromagnetic bath temperatures, again in very good agreement with experimental data.
Beyond these regimes, for an arbitrary quantum channel set in the far out-of-equilibrium situation where the temperature in the node significantly exceeds the one in the large electrodes, the equilibrium (uniform temperature) prediction for the conductance is recovered, albeit at a rescaled temperature $\alpha T_\mathrm{node}$.
\end{abstract}

\maketitle

The conductance of a quantum conductor embedded into an on-chip dissipative circuit is generally diminished at low temperatures and voltages.
This phenomenon originates from the granularity of charge transfers across non-ballistic quantum conductors.
The corresponding current shot noise excites the electromagnetic modes of the surrounding circuit, which suppresses the electrical conductance in proportion to the coupling to modes of unavailable high-energy (see \cite{SCT1992} for a review).
Taking into account this so-called dynamical Coulomb blockade (DCB) can be essential for the quantum nano-engineering of circuits assembled from several quantum components.
However, the many previous DCB studies mostly assumed a single, uniform temperature.
In contrast, driving composite nano-circuits usually involves an internal Joule dissipation, notably at the interconnect nodes (see e.g. \cite{Jezouin2013b}), which results in temperature gradients.
These gradients do not induce any thermoelectric currents in systems with a preserved electron-hole symmetry.
Nevertheless, they profoundly change the probability to excite the electromagnetic modes of the surrounding circuit, and therefore the DCB suppression of the electrical conductance.
Here, we investigate the DCB as a function of a quantitatively controlled temperature difference (bias) across a quantum conductor.
For this purpose, we focus more specifically on the revealing linear regime of small bias voltages.

The DCB was initially addressed theoretically and explored experimentally on tunnel junctions embedded into circuits described by a linear impedance \cite{Averin1986,Panyukov1988, Odintsov1988, Nazarov1989, Devoret1990, Girvin1990, Cleland1992, Holst1994, Joyez1998, Pierre2001, Hofheinz2011, Altimiras2014, Parlavecchio2015}.
In this limit, the junction can be treated as a small perturbation, thereby giving access to a full theoretical solution, including different temperatures for electrons on the large electrode side and the node side of the junction ($T$ and $T_\mathrm{node}$, resp.) and also for the electromagnetic modes of the linear impedance ($T_\mathrm{env}$).
The tunnel DCB theory is experimentally well established for arbitrary bias voltages and any uniform temperature ($T=T_\mathrm{node}=T_\mathrm{env}$).
This quantitative understanding allows one to exploit DCB as a tool, for example as a primary electron thermometer \cite{Iftikhar2016}.
In addition, the DCB across a tunnel junction in series with a relatively low resistance ($R\sim1.5\,\mathrm{k}\Omega\ll R_\mathrm{K}=h/e^2\simeq26\,\mathrm{k}\Omega$, with $h$ the Planck constant and $e$ the electron charge) was previously used as a probe for the unknown energy distribution of electrons driven out-of-equilibrium \cite{Anthore2003} (see \cite{Gutman2008} for a related theoretical development).
However, in that case, the validity of the tunnel DCB theory beyond a uniform temperature was assumed.
Here, in a first step, we put to experimental test the tunnel DCB theory in the presence of a thermal bias across the junction.

Second, beyond the limit of tunnel quantum conductors, the DCB theory itself remains very much incomplete (for notable theoretical and experimental advances, see e.g. \cite{Flensberg1993, Yeyati2001, Golubev2001, Cron2001, Kindermann2003, Safi2004, Golubev2005, Altimiras2007, Parmentier2011, Zamoum2012, Mebrahtu2012, Jezouin2013, Aristov2014, Altimiras2016, Anthore2018, Anthore2020, Boulat2020}). 
An important exception is for a single (spin-polarized) quantum channel of arbitrary electron transmission probability $\tau\in[0,1]$ set in series with a linear resistance $R$.
Under those circumstances, a fruitful mapping has been established with the problem of a spinless Luttinger liquid of interaction parameter $K=1/(1+R/R_\mathrm{K})$ with a single impurity \cite{Safi2004}.
Exploiting the results for a Luttinger liquid \cite{Kane1992b,Fendley1995b,Boulat2020}, this mapping provides the DCB conductance and all current cumulants at arbitrary voltages and any uniform temperature well below the capacitive cutoff ($eV,k_\mathrm{B}T\ll h/2\pi RC$).
Remarkably, a crossover toward an insulating state in the low temperature limit persists with any non-zero series resistance $R$, even for a channel that is almost (but not perfectly) ballistic above the DCB capacitive cutoff \cite{Safi2004, Zamoum2012}.
These theoretical predictions were found in precise quantitative agreement with experiments \cite{Mebrahtu2012, Jezouin2013, Anthore2018, Anthore2020} (as well as with numerical functional renormalization group simulations of a one dimensional electron lattice \cite{Anthore2020} and also approximate but remarkably accurate Luttinger liquid expressions \cite{Aristov2014}).
However, in the presence of a temperature bias across the quantum conductor and/or the linear series resistance, the DCB-Luttinger mapping breaks down and, consequently, so does the resulting DCB solutions.
The present work expands the DCB predictions to encompass near-ballistic quantum conductors submitted to an arbitrary temperature bias ($T$, $T_\mathrm{node}$), with a linear series resistance characterized by a third, `electromagnetic environment' temperature ($T_\mathrm{env}$), and establishes these new predictions experimentally.
In addition, beyond the tunnel and near-ballistic limits, 
we experimentally characterize at arbitrary transmission $\tau$ the deviations induced by a temperature bias, with respect to the conductance at a uniform temperature.
A simplification is found to occur for large temperature differences ($T_\mathrm{node}\gg T$), where the measured conductance approaches the uniform temperature prediction, although at a rescaled value of the temperature $\alpha T_\mathrm{node}$ significantly above the mean value ($\alpha>0.5$).
This finding generalizes a behavior that we specifically derive theoretically in the tunnel and near ballistic limits.

\begin{figure}
\centering\includegraphics[width=\columnwidth]{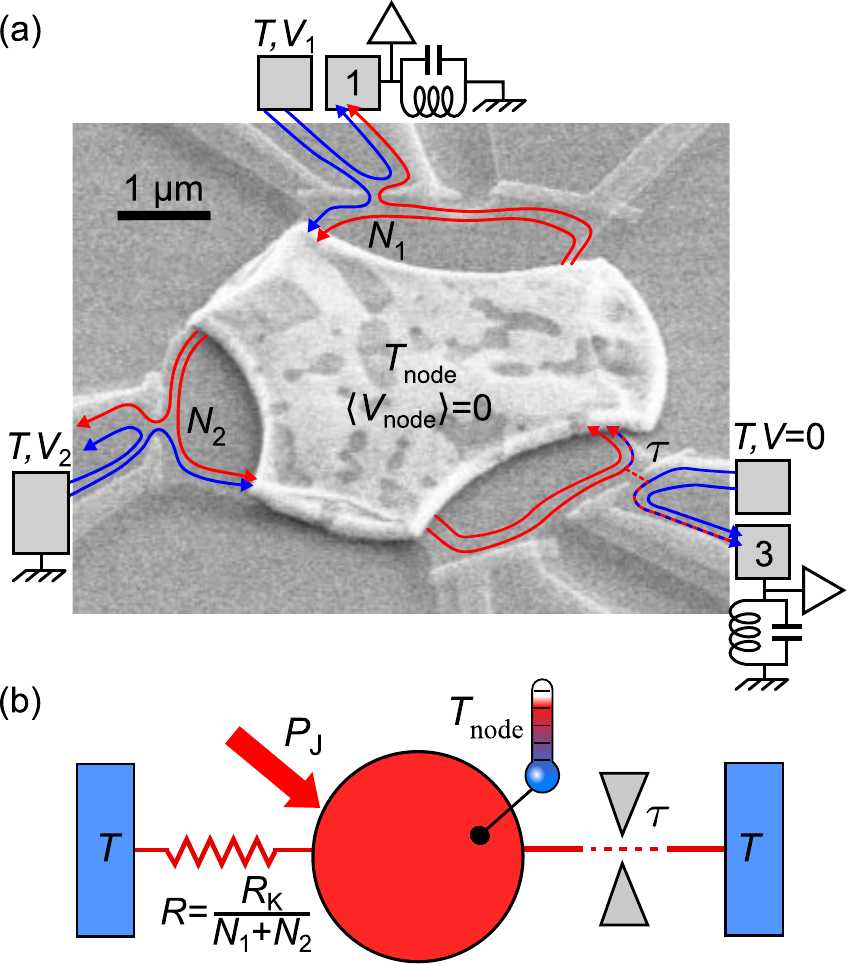}
\caption{(a) Device e-beam micrograph.
A single generic channel of electron transmission probability $\tau \in [0,1]$ (with $\tau/R_\mathrm{K}$ its conductance; $R_\mathrm{K}\equiv h/e^2$), as well as $N_1$ and $N_2$ fully transmitted channels ($N_1=N_2=1$ shown) are separately connected to a small metallic island (light gray). 
The two quantum Hall edge channels are represented as lines. 
A Joule heating of the island is realized by applying balanced voltages across the fully transmitted channels ($N_1 V_1 + N_2 V_2 =0$), such that the island's dc voltage $\left< V_\mathrm{node}\right>$ remains zero.
The heated-up island temperature $T_\mathrm{node}$ is monitored through noise measurements.
(b) Schematic representation.
The total number of fully transmitted channels controls the series resistance $R=R_\mathrm{K}/(N_1+N_2)$. 
The applied Joule power $P_\mathrm{J}$ results in a temperature bias $T_\mathrm{node}-T$ across the generic channel and also across $R$.
In contrast with the thermally biased generic channel, the $RC$ circuit is here approximately modeled as a heated up electromagnetic environment at a uniform temperature $T_\mathrm{env}$ generally taken as $(T+T_\mathrm{node})/2$.
}
\label{fig-sample}
\end{figure}

The studied quantum conductor (see Fig.~1(a)) consists in a single, fully tunable electronic channel realized by a quantum point contact (QPC) formed by field effect in a high-mobility Ga(Al)As 2D electron gas (2DEG) located 105\,nm below the surface.
The 2DEG is immersed in a large perpendicular magnetic field $B\simeq4$\,T, corresponding to the integer quantum Hall regime at filling factor $\nu=2$.
By lifting the spin degeneracy, the applied $B$ allows us to implement a single tunable channel.
This generic short channel (red dashed line) is connected on one side to a small floating circuit node at a temperature $T_\mathrm{node}$ (light gray central part and red disk in Fig.~1(a) and (b), respectively), and further away on the other side to a macroscopic electrode at a temperature $T$ (right rectangles in panels (a) and (b)).
The central node is also connected through two different paths to additional macroscopic electrodes at the same temperature $T$ (left and top rectangles in panel (a)), each path being composed of either one or two ballistic quantum Hall channels ($N_{1,2}\in\{1,2\}$).
These parallel ballistic channels implement altogether a precisely known linear resistance $R=R_\mathrm{K}/N$, with $N=N_1+N_2 \in \{2,3,4\}$, which is in series with the studied generic channel (see Appendix \ref{AppendixSample} for a discussion). 
As this resistance consists of $N$ chiral channels emitted from the heated node at $T_\mathrm{node}$ (red on Fig.~\ref{fig-sample}) and $N$ emitted from large ohmic contacts at $T$ (blue on Fig.~\ref{fig-sample}), it is also submitted to the same temperature bias as the quantum conductor.
In contrast, however, the electromagnetic modes of the corresponding $RC$ circuit are here modeled theoretically as being at a well defined, uniform electromagnetic environment temperature $T_\mathrm{env}$.
This approximation may be justified by the relatively low effect of $T_\mathrm{env}$ (see Appendix~\ref{AppendixTenv} for a discussion). 
In practice, unless stated otherwise, the plausible mean value $T_\mathrm{env}=(T+T_\mathrm{node})/2$ (see Fig.~\ref{fig-sample}(b)) is systematically used for the data/theory comparison (together with distinct $T$ and $T_\mathrm{node}$ across the studied channel).
The small floating node is effectively realized by a micron-size metallic island that is thermally diffused into the Ga(Al)As to make contact with the buried 2DEG.
The small size of the metallic island results in a small geometrical capacitance $C\simeq2.5$\,fF.  
(Additional measurements with $R=R_\mathrm{K}/2$ were performed on a different sample with $C\simeq3.1$\,fF.)
This is essential for a well-developed DCB effect, as the capacitance effectively short-circuits the series resistance for all possible energy exchanges with the electromagnetic environment that are comparable to or higher than the energy of one photon at the cutoff frequency $h/2\pi RC$.
In practice, for uniform temperatures ($T\simeq T_\mathrm{node}\simeq T_\mathrm{env}$) and low bias voltages ($eV\ll k_\mathrm{B}T$), the influence of the capacitance on the temperature dependence of the conductance becomes notable when $3 k_\mathrm{B}T\gtrsim h/2\pi RC$ ($T\gtrsim N\times30$\,mK for $C\sim3$\,fF and $R=R_\mathrm{K}/N$).

The floating node / metallic island is heated up by the locally dissipated Joule power $P_\mathrm{J}=(N_1V_1^2+N_2V_2^2)/2R_\mathrm{K}$ (note the factor $1/2$ \cite{Jezouin2013b}), with $V_{1(2)}$ the voltage feeding the path along $N_{1(2)}$ quantum Hall edge channels (see Fig.~1(a)).
In order to avoid the simultaneous build-up of a bias voltage across the studied generic channel, we apply balanced voltages of opposite signs $N_1V_1=-N_2V_2$, such that the node / island remains at zero dc voltage $\left< V_\mathrm{node} \right>\simeq0$.
Note that the dc thermoelectric current through the generic channel is directly measured and remains negligible (corresponding to $e| V_\mathrm{node}|<0.01 k_\mathrm{B} T_\mathrm{node}$), as expected from particle-hole symmetry (see e.g. \cite{ines_PRB_2019} for a  discussion). 
The resulting increase in the node temperature $T_\mathrm{node}$ depends on heat evacuation through each of the connected electronic channels and toward the phonons (see \cite{Sivre2019} for an experimental investigation involving one generic quantum channel).
In practice, the strong increase of the electron-phonon heat flow with temperature limits us to $T_\mathrm{node}\lesssim100$\,mK.

The temperature $T_\mathrm{node}$ can be experimentally determined from the electrical current fluctuations emitted from the metallic island \cite{Jezouin2013b, Sivre2018, Sivre2019}.
It should be noted that the measured noise involves the contributions of two sources of noise that both depend on $T_\mathrm{node}$: the thermal fluctuations of the current emitted from the metallic island, and the shot noise induced by the temperature difference across the studied non-ballistic quantum channel (also called $\delta T$-noise \cite{Lumbroso2018, Sivre2019, Larocque2020}).
Following the procedure established in \cite{Sivre2019}, we separate these two noise contributions by performing two independent noise measurements, on different electrodes.
Then we determine the increase in $T_\mathrm{node}$ solely from the thermal fluctuations (see Appendix~\ref{App-NoiseThermometry}).
Alternatively, it is also possible to calculate $T_\mathrm{node}$ based on the heat flow theory that was experimentally validated with a high accuracy in \cite{Sivre2019} on a similar device.
Here, the measured $T_\mathrm{node}$ are found to match calculated values within a negligible error $<4$\% (see Appendix Fig.~\ref{fig-siTcalcVersusTmeas}).
In the following, the displayed node temperatures $T_\mathrm{node}$ at base temperature $T\simeq8$\,mK are obtained from noise measurements, except for the limit case of tunnel junctions. 
In that tunnel case, and also for higher values of the large electrodes' temperature $T\gtrsim 15$\,mK, the calculated $T_\mathrm{node}$ were used (as performing noise measurements with sufficient resolution is time consuming).

\begin{figure}
\centering\includegraphics[width=\columnwidth]{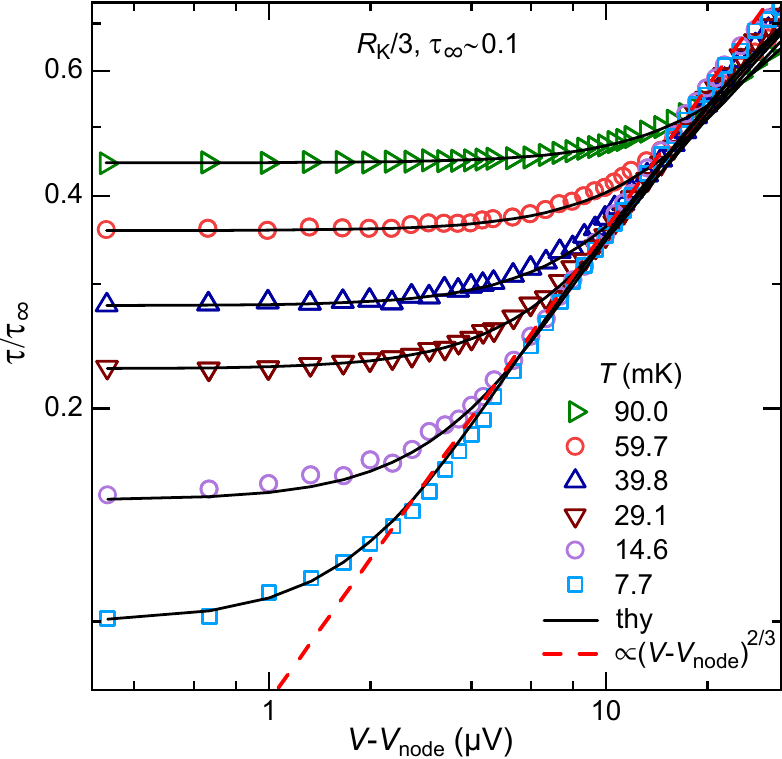}
\caption{Renormalized transmission probability versus bias voltage in the tunnel regime.
Symbols represent, in a log-log scale versus $V-V_\mathrm{node}$ and for different uniform temperatures $T\simeq T_\mathrm{node}\simeq T_\mathrm{env}$, the experimental values of $\tau/\tau_\infty$. 
Those are obtained from $R_\mathrm{K}/\tau$=1/$G-R$, with $G$ the measured differential conductance of the sample including both tunnel contact and series resistance $R=R_\mathrm{K}/3$ (see Appendix Fig.~\ref{fig-SI-DCBTunnel-V} for similar data at $R=R_\mathrm{K}/2$ and $R_\mathrm{K}/4$).
The intrinsic (unrenormalized) transmission probability $\tau_\infty$ depends on applied gate voltages, and constitutes the only adjustable parameter in the data-theory comparison.
Black continuous lines: full quantitative predictions of the tunnel DCB theory (see Eq.~\eqref{eqTunnelP(E)} and Appendix~\ref{DCBtunnel}).
Red dashed line: power law $\eta \left(V-V_\mathrm{node}\right)^{2R/R_\mathrm{K}}$ predicted for $k_\mathrm{B}T\ll e(V-V_\mathrm{node})\ll h/2\pi RC$, using the quantitative theoretical value of $\eta$ (see Appendix Eq.~\eqref{TunnelVtoZero}).}
\label{fig-dcbTunnelV}
\end{figure}

We first focus on the tunnel limit of a quantum channel of small intrinsic (not renormalized by DCB) transmission probability $\tau_\infty\ll1$.
This regime is described by the tunnel DCB theory, also called $P(\epsilon)$ theory \cite{SCT1992}.
In this framework, the transmission probability $\tau$ reduced by DCB (i.e.\ the channel's differential conductance in units of $e^2/h$) can be expressed as a function of the Fermi distributions $f_\mathrm{T(T_\mathrm{node})}(E)$ in the electrodes at the temperatures $T$ and $T_\mathrm{node}$ on either side of the tunnel contact, as well as of the probability $P_\mathrm{T_\mathrm{env}}(\epsilon)$ that the energy $\epsilon$ is absorbed ($\epsilon>0$) or emitted ($\epsilon<0$) by the electromagnetic environment at a temperature $T_\mathrm{env}$:
\begin{equation}
\begin{split}
    \tau/\tau_\infty=1+
    \int & d\epsilon dE\, P_\mathrm{T_\mathrm{env}}(\epsilon) f_\mathrm{T}\left(E-e\left(V-V_\mathrm{node}\right)\right)\\
    &\left\{ \partial_\mathrm{E} f_\mathrm{T_\mathrm{node}}(E+\epsilon)-\partial_\mathrm{E} f_\mathrm{T_\mathrm{node}}(E-\epsilon)\right\},
\end{split}
\label{eqTunnelP(E)}
\end{equation}
where $V$ is the bias voltage applied to the electrode behind the generic channel (see Appendix~\ref{DCB} for full details and analytical asymptotic solutions).
We start by verifying the canonical DCB behavior of the tunnel quantum channel in the well-established regime of a uniform temperature $T_\mathrm{node}\simeq T\simeq T_\mathrm{env}$ as a function of $V$.
In practice, the transmission probability renormalized by DCB is obtained from $\tau =R_\mathrm{K}/(1/G-R)$, where $G$ is the measured differential conductance across the whole sample (generic channel in series with $R$).
Note that $\tau_\infty$ is not accurately known experimentally (approximately obtained from the channel's conductance in $e^2/h$ units at large voltage bias where the DCB renormalization is small).
Therefore, unless stated otherwise, $\tau_\infty$ is considered as an adjustable parameter applying globally to all the measurements performed with the same setting of the device.
As shown in Fig.~\ref{fig-dcbTunnelV} for $R=R_\mathrm{K}/3$, a good agreement is observed between data and theoretical predictions without any other adjustable parameters (see also Appendix Fig.~\ref{fig-SI-DCBTunnel-V} for $R=R_\mathrm{K}/2$ and $R_\mathrm{K}/4$).
Henceforth, we focus on the conductance in the linear regime ($V \to 0$).

\begin{figure}[h!]
\centering\includegraphics[width=\columnwidth]{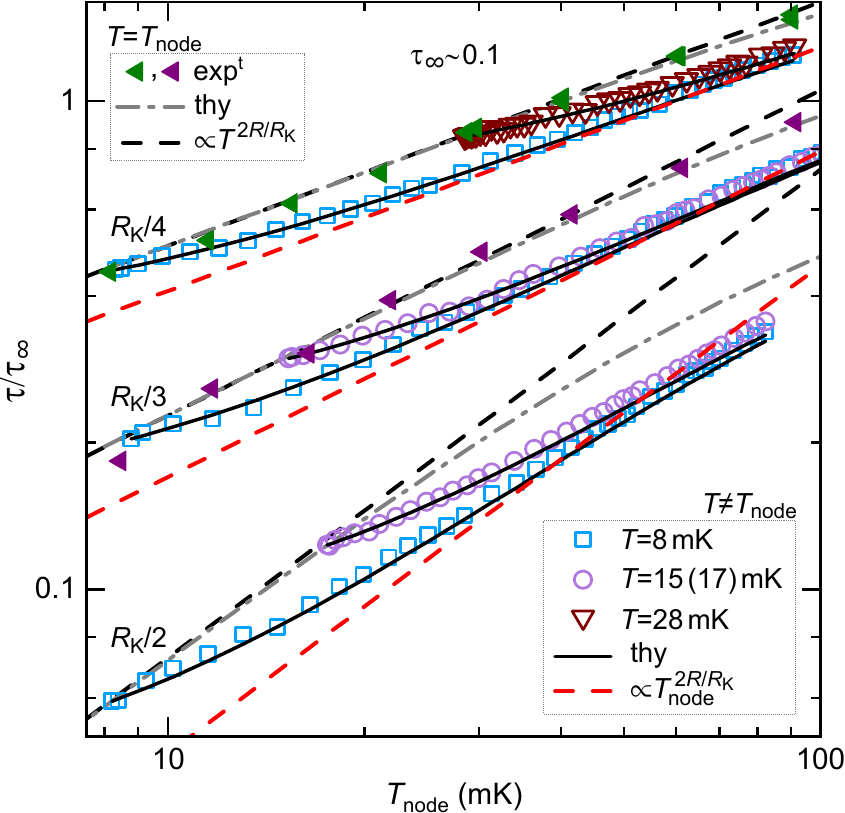}
\caption{Renormalized transmission probability versus node temperature in the tunnel regime ($\tau_\infty\ll1$) at low bias voltage ($V\to0$).
Full symbols: experimental $\tau/\tau_\infty$ at a uniform temperature ($T_\mathrm{node}\simeq T\simeq T_\mathrm{env}$) for $R=R_\mathrm{K}/3$ (purple) and $R_\mathrm{K}/4$ (green).
Gray dash-dotted lines and black dashed lines: full and asymptotic ($k_\mathrm{B}T\ll h/2\pi RC$) predictions, respectively, both of them for a uniform temperature $T=T_\mathrm{node}=T_\mathrm{env}$.
Open symbols: renormalized transmission probability data obtained in the presence of the series resistances $R=R_\mathrm{K}/2$, $R_\mathrm{K}/3$ and $R_\mathrm{K}/4$, with a temperature bias $T_\mathrm{node}-T\geq0$ developing from different fixed values of the temperature $T$.
Black continuous lines and red dashed lines: full and asymptotic ($k_\mathrm{B}T\ll k_\mathrm{B}T_\mathrm{node}\ll h/2\pi RC$) predictions, respectively, both of them versus $T_\mathrm{node}$ at fixed $T$ and assuming an electromagnetic environment at $T_\mathrm{env}=(T+T_\mathrm{node})/2$.
For clarity, the data/theory are shifted vertically in this log-log plot by a factor of 2 (4) for $R=R_\mathrm{K}/3$ ($R_\mathrm{K}/4$).
}
\label{fig-DCBtunnelT}
\end{figure}

In Fig.~\ref{fig-DCBtunnelT}, the transmission probability ratio $\tau/\tau_\infty$ is now shown versus node temperature $T_\mathrm{node}$ and in the presence of a series resistance $R=R_\mathrm{K}/N$ with $N\in\{2,3,4\}$.
For a uniform temperature ($T_\mathrm{node}\simeq T\simeq T_\mathrm{env}$), the data points displayed as full symbols follow the power law $T_\mathrm{node}^{2R/R_\mathrm{K}}$ (black dashed lines) predicted at $k_\mathrm{B}T_\mathrm{node}\ll h/2\pi RC$ (see Appendix Eq.~\eqref{LowTeqPowerLaw} for a novel exact analytical expression).
Significant deviations from the asymptotic power law are only expected to show up for the largest uniform temperatures achieved at $R=R_\mathrm{K}/2$.
This can be seen in Fig.~\ref{fig-DCBtunnelT} by comparing the black dashed lines with the full predictions of the tunnel DCB theory displayed as dash-dotted gray lines (see Appendix~\ref{DCBtunnel}, and also Appendix Eqs.~\eqref{Geq2}, \eqref{eqJdagger} and \eqref{eqJdaggerMatsubara} for a numerically more efficient formulation of the linear conductance).
With this precise confirmation of the canonical tunnel DCB behavior of our device versus uniform temperature, we can now investigate the influence of a temperature bias.

Measurements performed for a fixed $T$ of either 8\,mK, 15\,mK or 28\,mK, as a function of the heated-up $T_\mathrm{node}$, are shown in Fig.~\ref{fig-DCBtunnelT} as open symbols.
The transmission probability directly separates itself from the power law predicted at low uniform temperatures (black dashed lines).
Yet, we find that the same power law $T^{2R/R_\mathrm{K}}$ is recovered for large $T_\mathrm{node}/T\gtrsim4$, although with a lower multiplicative factor (red dashed lines).
This behavior is expected from the tunnel DCB theory.
We derived a novel exact analytical expression, given in Appendix Eq.~\eqref{LowToneNulPowerLaw}, for the conductance in the limit of large temperature bias across the tunnel junction ($T_\mathrm{node}\gg T$), assuming the electromagnetic environment is at the average temperature $T_\mathrm{env}\simeq T_\mathrm{node}/2$ (see part (\textit{ii}) of Appendix~\ref{DCBtunnelTV0} for a derivation).
The red dashed lines were obtained using this expression without adjustable parameters.
Compared to the power law at a uniform temperature $T_\mathrm{node}=T=T_\mathrm{env}$, we find here the same power law exponent of $T_\mathrm{node}$ but with a multiplicative factor reduced by $(2^{1-2R/R_\mathrm{K}}/\sqrt{\pi})\Gamma(1.5+R/R_\mathrm{K})/\Gamma(1+R/R_\mathrm{K})$ ($\Gamma$ is the gamma function), in quantitative agreement with experimental observations.
It is useful to note that the above mentioned reduction factor on the conductance can be formulated as a rescaling in temperature by a reduction factor $\alpha$ (as we are in the presence of a power law).
As discussed later, such a formulation is better suited for extrapolation beyond the tunnel regime.
Remarkably, this $T_\mathrm{node}$ rescaling factor of approximately $\alpha=0.637$, 0.648 and 0.655 for $R=R_\mathrm{K}/2,$ $R_\mathrm{K}/3$ and $R_\mathrm{K}/4$, respectively (see Appendix \ref{Appendix3T}), is markedly higher ($\sim+30\%$) than the $1/2$ factor corresponding to the asymptotic mean temperature $T_\mathrm{node}/2$.
The tunnel DCB theory also allows us to numerically evaluate $\tau/\tau_\infty$ over the complete span of $T_\mathrm{node}\geq T$.
The black continuous lines show such calculations performed assuming $T_\mathrm{env}=(T+T_\mathrm{node})/2$.
(Note that our experimental accuracy does not allow us to precisely resolve what is the most appropriate choice for $T_\mathrm{env}$ in our device, with a similarly good agreement at $T_\mathrm{env}\in[(T+T_\mathrm{node})/2,T_\mathrm{node}]$, see Appendix~\ref{AppendixTenv} and Appendix Fig.~\ref{fig-SI-DCBTunnel-Tenv}). 
As previously, the only adjustable parameter in the data/theory comparison is $\tau_\infty$, which is here set by matching the theory at a uniform temperature with the lowest $T_\mathrm{node}$ data point (where $T_\mathrm{node}\simeq T$).
The precision agreement between data and full numerical calculations observed in Fig.~\ref{fig-DCBtunnelT} establishes the validity of the tunnel DCB theory of a temperature biased tunnel junction.

We now investigate the more general case of an electronic quantum channel of arbitrary electron transmission probability ($\tau_\infty \in[0,1]$) submitted to a temperature bias.
Theoretically, arbitrary values of $\tau_\infty$ can be addressed through the mapping of the DCB problem in the presence of a series resistance $R$ to that of one impurity in a Luttinger liquid \cite{Safi2004}, which holds for uniform temperatures $T\simeq T_\mathrm{node}\simeq T_\mathrm{env}$ and arbitrary voltages $V$ provided that $eV,k_\mathrm{B}T\ll h/2\pi RC$.
The corresponding Luttinger crossover toward an insulating state at low temperatures and voltages was fully solved first for special values of the resistance in the pioneer work \cite{Fendley1992} (reducing to a simple analytical expression for $R=R_\mathrm{K}$, see e.g.\ \cite{Zamoum2012}) and recently for any value of $R$ \cite{Boulat2020}.
This solution takes, in the linear regime $V\to 0$, the form of a renormalization curve $G^\mathrm{eq}_{R/R_\mathrm{K}}(T/T_\mathrm{I})$ for the sample conductance. 
It is said `universal' because the influence of microscopic parameters, such as the high-energy capacitive cutoff and the intrinsic channel transmission $\tau_\infty$ in the absence of DCB, are encapsulated into the single temperature scale $T_\mathrm{I}$.
The previously discussed tunnel regime corresponds to the power-law expected when approaching the insulating low temperature limit $G^\mathrm{eq}_{R/R_\mathrm{K}}(T/T_\mathrm{I}\rightarrow 0)\propto (T/T_\mathrm{I})^{2R/R_\mathrm{K}}$.

\begin{figure}[h!]
\centering\includegraphics[width=\columnwidth]{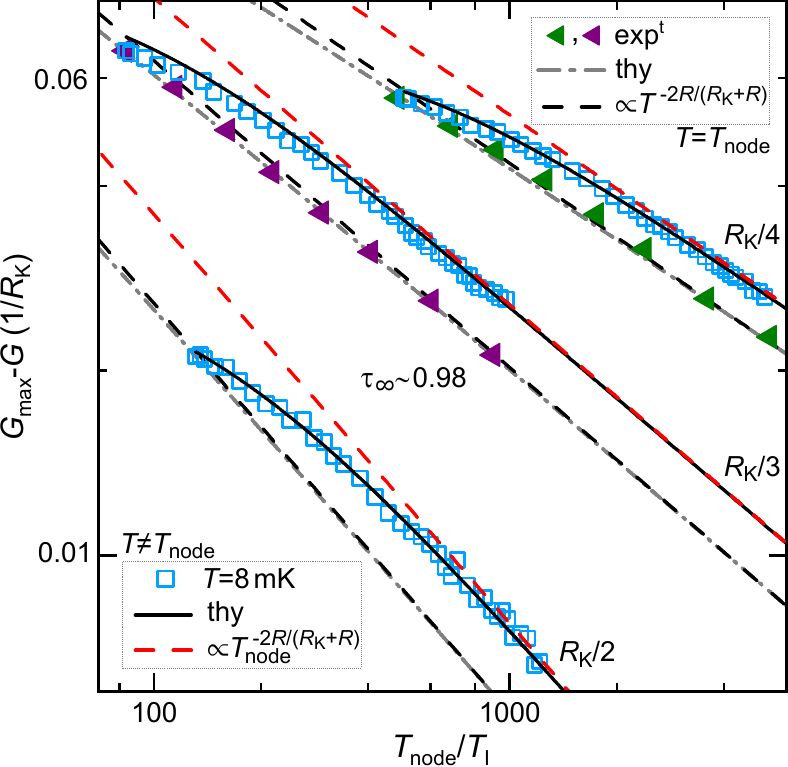}
\caption{Conductance versus node temperature in the near ballistic regime ($1-\tau\ll1$).
The difference $G_\mathrm{max}-G$, between ballistic limit $G_\mathrm{max}=\frac{N}{N+1}R_\mathrm{K}^{-1}$ and sample linear conductance $G(V\to0)$, is plotted in a log-log scale as a function of the rescaled node temperature $T_\mathrm{node}/T_\mathrm{I}$ for $R=R_\mathrm{K}/2,$ $R_\mathrm{K}/3$ and $R_\mathrm{K}/4$.
Gray dash-dotted lines: full universal predictions at uniform temperatures ($T_\mathrm{node}=T=T_\mathrm{env}$) \cite{Boulat2020}.
Black and red dashed lines: asymptotic ($G_\mathrm{max}-G\rightarrow0$) power law predictions at uniform temperatures (black) and in the limit $T_\mathrm{node}\gg T$ with $T_\mathrm{env}=T_\mathrm{node}/2$ (red).
Black continuous lines: predictions for a near ballistic junction temperature biased at $T_\mathrm{node}\geq T$ with $T=8\,$mK and $T_\mathrm{env}=(T+T_\mathrm{node})/2$.
Full and open symbols: measurements at a uniform temperature $T\simeq T_\mathrm{node}$ (full) and in the presence of a temperature bias at fixed $T\simeq8$\,mK (open), respectively.
The scaling temperature $T_\mathrm{I}$ is set by adjusting to the universal theory the lowest temperature point $T_\mathrm{node}\simeq 8$\,mK.
}
\label{fig-tauClose1T}
\end{figure}

In a first step, let us consider the near ballistic regime ($1-\tau \ll 1$).
For a Luttinger liquid with an impurity, a duality is predicted between strong back-scattering (tunnel regime) with a Luttinger interaction parameter $K$, and weak back-scattering (near ballistic regime) with an interaction parameter $1/K$ \cite{Kane1992b,Fendley1998,Lesage1999}. 
Thus the dual of the tunnel conductance power law $T^{2/K-2}$ reads in the near ballistic regime $G_\mathrm{max}-G\propto (T/T_\mathrm{I})^{2K-2}\propto (T/T_\mathrm{I})^{-2R/(R_\mathrm{K}+R)}$ with $G_\mathrm{max}=\frac{N}{N+1}R_\mathrm{K}^{-1}$ \cite{Safi2004}.
Accordingly, as shown in Fig.~\ref{fig-tauClose1T} for $R=R_\mathrm{K}/N$ with $N\in\{2,3,4\}$, the Luttinger universal renormalization curves at equilibrium $G^\mathrm{eq}_{R/R_\mathrm{K}}(T/T_\mathrm{I})$ (dash-dotted gray lines) asymptotically approach the corresponding power laws (black dashed lines) as $G_\mathrm{max}-G\to0$ ($T_\mathrm{node}/T_\mathrm{I}\to\infty$). 
However, this duality is not expected to hold beyond the regime of both low and uniform temperatures where the DCB-Luttinger liquid mapping applies.
Consequently, it is not expected to give access to the quantitative multiplicative factor for the conductance, as it depends on the capacitive cutoff.
Here, we overcome these limitations by providing new DCB predictions for near ballistic junctions ($G_\mathrm{max}-G\ll1$).
As detailed in Appendix~\ref{CMballistic}, we obtain a novel exact analytical expression for the conductance at low uniform temperatures ($T_\mathrm{node}=T=T_\mathrm{env}\ll h/2\pi k_\mathrm{B}RC$, see Appendix Eq.~\eqref{eqGmaxMinusGNearBallistic}), by expanding upon the approach of \cite{Furusaki1995b}.
Remarkably, the duality tunnel - near ballistic regime is found to hold not only for the power law exponent, but also for the numerical multiplicative factor.
In the presence of a temperature biased channel ($T\neq T_\mathrm{node}\neq T_\mathrm{env}\ll h/2\pi k_\mathrm{B}RC$), novel predictions are obtained within a different approach based on the Keldysh formalism  \cite{Safi_preparation}, and take the form of an integral readily evaluated numerically (see Appendix~\ref{ISballistic3T}).
The black continuous lines in Fig.~\ref{fig-tauClose1T} display these new predictions, obtained by assuming that the electromagnetic environment is at the mean temperature $T_\mathrm{env}=(T+T_\mathrm{node})/2$ (see Appendix Eq.~\eqref{eqDTweakbs}).
Similarly to the tunnel regime, we predict that the power law dependence of $G_\mathrm{max}-G$ for uniform temperatures is recovered at $T_\mathrm{node}\gg T$, with a change in the multiplicative factor displayed by the shift in log-log scale between black and red dashed lines.
This factor change here applies to the difference $G_\mathrm{max}-G$, instead of the channel's conductance in the opposite tunnel regime. 
For a quantitative comparison between these two regimes, the factor change can also be recast as an effective reduction of the node temperature $T_\mathrm{node}\rightarrow \alpha T_\mathrm{node}$.
We find $\alpha\simeq 0.61$, $0.62$ and $0.64$ for $R=R_\mathrm{K}/2$, $R_\mathrm{K}/3$ and $R_\mathrm{K}/4$, respectively.
Although not exactly identical, the temperature reduction factor $\alpha$ in the tunnel and quasi-ballistic regimes are very close to one another, within $5\%$ for each of the presently investigated series resistances.
However, larger differences between these regimes are predicted to develop for larger series resistances.

Comparing theory and experiment in the near ballistic regime, first for uniform temperatures  ($T_\mathrm{node}\simeq T\simeq T_\mathrm{env}$), we find a very good agreement between the full universal theory curves (dash-dotted gray lines) and the data (full symbols) for $R=R_\mathrm{K}/3$ and $R_\mathrm{K}/4$. 
The only fit parameter is the value of $T_\mathrm{I}$, determined by matching the data point at the lowest temperature with the predicted value  $G^\mathrm{eq}_{R/R_\mathrm{K}}(T_\mathrm{node}/T_\mathrm{I})$ (see Appendix~\ref{ballisticCMvsData} and Appendix Fig.~\ref{fig-SI-Close1PowerLaw} for a parameter-free comparison, limited by the experimental uncertainty on $\tau_\infty$). 
Note that a good match with the corresponding asymptotic power law (black dashed lines) is observed up to the highest uniform temperature of about 100\,mK.
Second, the node temperature is now changed while $T\simeq8$\,mK is kept fixed to create a temperature bias.
The corresponding $G_\mathrm{max}-G$ data points (blue open symbols) depart from the predictions at a uniform temperature $T_\mathrm{node}=T=T_\mathrm{env}$ (dash-dotted gray lines and black dashed lines). 
In contrast, a precise agreement is observed with the presently developed DCB predictions for a temperature-biased near-ballistic channel, here evaluated at fixed $T=8\,$mK and assuming $T_\mathrm{env}=(T+T_\mathrm{node})/2$ (black continuous lines).

\begin{figure}
\centering\includegraphics[width=\columnwidth]{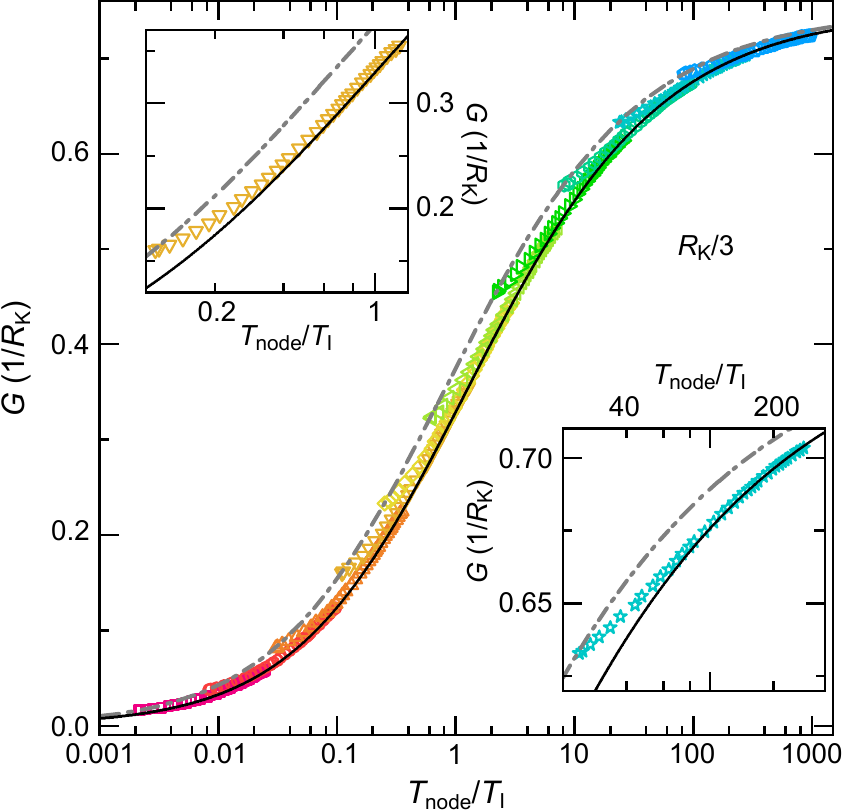}
\caption{Conductance for arbitrary channel tunings and $R=R_\mathrm{K}/3$ versus node temperature.
Open symbols: measured sample conductance $G$ at $T\simeq8\,$mK versus heated-up $T_\mathrm{node}$.
Different tunings ($\tau_\infty$) of the channel are shown using different symbols and colors.
For each tuning a unique value of $T_\mathrm{I}$ is determined by matching the lowest temperature data point at $T_\mathrm{node}\simeq T\simeq8$\,mK with $G^\mathrm{eq}_{1/3}(T/T_\mathrm{I})$, the universal prediction at uniform temperatures (gray dash-dotted line).
The black line corresponds to the universal prediction with the temperature reduction factor $T_\mathrm{node}\rightarrow\alpha T_\mathrm{node}$ predicted to apply for large temperature bias in the tunnel regime ($\alpha\simeq0.648$, see Appendix Eq.~\eqref{eq-rescaledTI}).
Insets: magnified view for intermediate channel tunings.
}
\label{fig-GeneralTau}
\end{figure}

Does an effective temperature rescaling at $T_\mathrm{node}\gg T$ persist for an arbitrary transmission probability $\tau\in[0,1]$ across the generic quantum channel, beyond the tunnel and near ballistic limits?
First, as previously demonstrated \cite{Anthore2018, Anthore2020}, we find a good agreement over the full range of conductance $0<G<G_\mathrm{max}$ between the universal renormalization curves at uniform temperatures $G^\mathrm{eq}_{R/R_\mathrm{K}}(T_\mathrm{node}/T_\mathrm{I})$ (gray dash-dotted lines in Fig.~\ref{fig-GeneralTau} and Appendix Fig.~\ref{fig-SI-FullRKs2RKs4}) and the sample conductance measured at $T_\mathrm{node}\simeq T$ (see Appendix Fig.~\ref{fig-SI-HomodGsdLnT} for a comparison).
Now turning to the regime of a temperature bias $T_\mathrm{node}\geq T$, we focus in Fig.~\ref{fig-GeneralTau} on the representative $R=R_\mathrm{K}/3$ at $T\simeq8$\,mK for clarity (see Appendix Fig.~\ref{fig-SI-FullRKs2RKs4} for $R=R_\mathrm{K}/2$ and $R_\mathrm{K}/4$).
The measured conductance $G$ across the whole sample -- non-ballistic channel and $R$ -- is displayed as open symbols, with each identical set of symbols of the same color corresponding to a fixed tuning of the quantum channel (a fixed $\tau_\infty$, see insets in Fig.~\ref{fig-GeneralTau} for two data sets each corresponding to a different tuning).
A single value of the scaling temperature $T_\mathrm{I}$ is associated with a fixed tuning of the channel, together with a specific $RC$ environment. 
It is here obtained by matching the equilibrium $G^\mathrm{eq}_{1/3}(T_\mathrm{node}/T_\mathrm{I})$ with the conductance measured for this tuning at the lowest $T_\mathrm{node}$ where the temperature is uniform ($T_\mathrm{node}\simeq T\simeq8$\,mK).
As $T_\mathrm{node}$ increases, a temperature bias develops and the measured conductance moves away from $G^\mathrm{eq}_{1/3}(T_\mathrm{node}/T_\mathrm{I})$.
Remarkably, we find that for large enough temperature bias ($T_\mathrm{node}/T\gtrsim4$), the universal uniform temperature curve is recovered at the experimental resolution provided we apply a temperature rescaling $T_\mathrm{node}\rightarrow\alpha T_\mathrm{node}$.
The temperature reduction factor in the tunnel and near ballistic
regimes being very close to one another (within 5\%), we only display for this comparison $G^\mathrm{eq}_{1/3}(\alpha T_\mathrm{node}/T_\mathrm{I})$ with $\alpha$ the tunnel temperature reduction factor given by Appendix Eq.~\eqref{eq-rescaledTI} (black continuous line).

The conductance reduction experienced by a quantum conductor when it is embedded into an on-chip dissipative circuit, the so-called dynamical Coulomb blockade (DCB), has been explored in the presence of temperature gradients. 
We experimentally have established the existing tunnel DCB theory under a temperature bias, and obtained novel analytical expressions for the conductance.
In the near ballistic regime, we have developed the theory and provided quantitative predictions as a function of the device parameters and for arbitrary temperatures differences, which have been verified experimentally.
More generally, beyond the tunnel and near ballistic limits, we have observed that the equilibrium predictions apply to a good approximation for large temperature differences, provided a simple effective rescaling of the temperature is performed. 
This work develops and establishes our understanding of thermally inhomogeneous quantum circuits, a knowledge set to play a role for the future engineering of functional quantum devices involving local dissipation. 

%

\begin{acknowledgments}
This work was supported by the French RENATECH network, the national French program `Investissements d'Avenir' (Labex NanoSaclay, ANR-10-LABX-0035) and the French National Research Agency (projects QuTherm, ANR-16-CE30-0010-01 and SIM-CIRCUIT, ANR-18-CE47-0014-01).

E.S., F.P. and H.D. performed the experiment with inputs from A.Aa. and A.An.;
A.An., E.S., F.P. and H.D. analyzed the data;
A.An. and F.D.P. fabricated the sample;
A.C., A.O. and U.G. grew the 2DEG;
A.An. and C.M. extended the tunnel DCB theory.
C.M. obtained quantitative equilibrium predictions in the near ballistic regime;
I.S. developed the theory for different bath temperatures in the near ballistic regime;
F.P. wrote the paper with inputs from all authors;
A.An. wrote the appendix with F.P., except theory sections in near ballistic regime by I.S. and C.M., with inputs from all authors;
A.An. and F.P. led the project.
\end{acknowledgments}

\section*{Appendix} 

\subsection{Samples}
\label{AppendixSample}
The sample was made in the same batch as the one used in \cite{Iftikhar2016}, with additional fabrication steps.
It consists of a Ga(Al)As two-dimensional electron gas buried $105$\,nm below the surface, of density $2.5 \times 10^{11}\,\mathrm{cm}^{-2}$ and of mobility $10^6\,\mathrm{cm}^2\mathrm{V}^{-1}\mathrm{s}^{-1}$. 
Its nanostructuration is performed by standard e-beam lithography, dry etching and metallic deposition. 
The  central metallic island (nickel (30\,nm), gold (120\,nm) and germanium (60\,nm)) was thermally annealed (440\,$^\circ$C for 50\,s) to achieve a good ohmic contact with the 2DEG.\\
The contact quality between the metallic island and the 2DEG is fully characterized, through the individual determination of the electron reflection probability at the interface for each connected quantum Hall channel, with the same experimental procedure previously detailed in Methods of \cite{Iftikhar2015}.
We find a reflection probability below $\lesssim0.001\%$ (the statistical uncertainty) for all the used channels.\\
The typical electronic level spacing in the metallic island is estimated to be negligibly small ($\delta\approx k_\mathrm{B}\times0.2\,\mu$K), based on the electronic density of states of gold ($\nu_\mathrm{F}\approx1.14\times 10^{47}$\,J$^{-1}$m$^{-3}$) and the metallic island volume ($\approx3\,\mu$m$^3$).\\
Due to technical problems, the initial electrostatic gates (shown on Fig.~\ref{fig-sample}) were etched out, redefined and redeposited with $40$\,nm of aluminum.\\
An important device parameter is the charging energy $E_\mathrm{C}\equiv e^2/2C$ of the island.
For this specific device, one of the channel remains open and standard Coulomb diamond determination of $E_\mathrm{C}$ could not be performed.
Instead, the value $E_\mathrm{C}\simeq k_\mathrm{B}\times 0.37$\,K is obtained by fitting the overall $G(V)$ tunnel data at a uniform temperature, at all $T\simeq T_\mathrm{node}\simeq T_\mathrm{env}$ (from $8$ to $90$\,mK) and for all the series resistances $R_\mathrm{K}/N$ with $N\in\{2,3,4\}$.
Note that this value is slightly higher ($\sim+20\%$) than the one found in \cite{Iftikhar2016}.
The reduced geometrical capacitance $C$ might come from imperfect new gates.  
However, note also that $E_\mathrm{C}$ is sufficiently large to have a relatively small impact, notably in the data-theory comparison as a function of a temperature bias.
This can be seen in Fig.~\ref{fig-DCBtunnelT} from the small difference between black continuous lines and red dashed lines at the largest $T_\mathrm{node}$.\\
The resistance in series with the studied, non-ballistic channels is simply taken as the dc electrical resistance $R_\mathrm{K}/N$ of the constitutive $N$ quantum Hall channels in parallel.
In principle, deviations from this value could occur at high frequencies.
However, these deviations are limited by the high-frequency cutoff introduced by the capacitance $C$ of the island.
For instance, such deviations could result from a non-zero conductivity across the 2D bulk at the frequency $\nu$ when the energy $h\nu$ becomes comparable with the quantum Hall gap.
Yet, in our sample the quantum Hall gap is about two orders of magnitude higher than $h/RC=N\times E_\mathrm{C}$.
Also, the transit time $t_\mathrm{transit}$ along the micron-scale distance between the back-scattering location in the non-ballistic channel and the island could lead to an inductive correction to $R$.
Yet, given the typical velocity of $\sim 10^5\,$m/s for the propagation of charge along the quantum Hall edge, the associated energy scale $h/t_\mathrm{transit}$ is about one order of magnitude higher than $h/RC$.
In practice, we check the validity of our $RC$ circuit description by comparing the conductance data with the DCB theory in the well-established regimes of a uniform temperature (see Figs.~2, 9, 10, and full symbols in Figs.~3 and 4).

\subsection{$P(E)$ theory of dynamical Coulomb Blockade for a tunnel junction}
\label{DCB}

Here we focus on the predictions for the DCB renormalization of the transmission probability $\tau$ across an electronic channel in the tunnel regime ($\tau,\tau_\infty\ll1$), when it is embedded in an $RC$ circuit.
These results can be applied to a high-resistance tunnel junction including many such channels replacing $\tau/R_\mathrm{K}$ and $\tau_\infty/R_\mathrm{K}$ by, respectively, the junction renormalized and intrinsic differential conductance.

\subsubsection{Numerically efficient formulation at arbitrary voltage and uniform temperature $T=T_\mathrm{node}=T_\mathrm{env}$ with an $RC$ environment}
\label{DCBtunnel}

Numerical calculations of the conductance of a coherent conductor in the tunnel limit in presence of environmental back-action were made with the efficient formulation of the DCB theory for small tunnel junctions given in \cite{Joyez1997}. 
In this section, we recapitulate the expressions used in the case of uniform temperatures and finite bias voltages.

The transmission probability $\tau\equiv R_\mathrm{K}\mathrm{d}I/\mathrm{d}V$ across a short electronic channel in the tunnel regime, embedded in an electromagnetic environment described by the series impedance $Z(\omega)$, at a uniform temperature $T=T_\mathrm{node}=T_\mathrm{env}$, and for a voltage $V$ applied here across the channel (corresponding to $V-V_\mathrm{node}$ in Fig.~\ref{fig-dcbTunnelV}, as shown in Fig.~\ref{fig-sample}(a)) reads \cite{Joyez1998}:
\begin{equation}
\begin{split}
\tau (&V,T)/\tau_\infty-1 = \\
&\int_0^{+\infty}\!dt\,\frac{2 \pi t}{\sinh^{2} \frac{\pi t k_\mathrm{B} T}{\hbar}}\left(\frac{k_\mathrm{B} T}{\hbar}\right)^2  \mathrm{Im}\left[e^{J(t)}\right] \cos \frac{e V t}{\hbar},
\label{PhilEquilibrium}
\end{split}
\end{equation}
with the channel intrinsic resistance $R_\mathrm{K}/\tau_\infty$ assumed to be very large compared to the environmental impedance $\mathrm{Re}[Z(\omega)] \ll R_\mathrm{K}/\tau_\infty$.

For the simplified RC model of the electromagnetic environment shown in article Fig.~\ref{fig-sample}(b) ($Z(\omega)=R/(1+iRC\omega)$), $J(t)$ reads:
\begin{equation}
\begin{split}
J(t)=\frac{\pi R}{R_\mathrm{K}}\Bigg(& \left(1-e^{- \mid t\mid/RC} \right) \left( \mathrm{cot}\frac{\hbar}{2RCk_\mathrm{B}T}-i \right)\\
&-\frac{2  k_\mathrm{B} T \mid t \mid}{\hbar} 
+ 2 \sum_{n=1}^{+\infty}\frac{ 1-e^{-\omega_n \mid t\mid}}{ \pi n \left[ (RC\omega_n)^2-1\right]}   \Bigg),
\end{split}
\label{eqJ(t)}
\end{equation}
where $\omega_n=2\pi n k_\mathrm{B} T/\hbar$ are Matsubara's frequencies and
\begin{equation}
\begin{split}
2 \sum_{n=1}^{+\infty}\frac{ 1-e^{-\omega_n  t}}{ \pi n \left[ (RC\omega_n)^2-1\right]} = - \frac{1}{\pi}\big[2\gamma +\Psi(-x)+\Psi(x)\\
+2\ln(1-y) + \frac{y}{1+x}~_2 F_1(1,1+x,2+x,y)
\\+ \frac{y}{1-x} ~_2F_1(1,1-x,2-x,y) \big],
\end{split}
\end{equation}
where $\gamma\simeq0.5772$ is Euler's constant, $\Psi$ is the logarithmic derivative of the Gamma function, $_2F_1$ is the hypergeometric
function, $y=\mathrm{exp}(\frac{-2 \pi t k\mathrm{_B} T}{\hbar})$, and $x=E_\mathrm{C} R_\mathrm{K}/(2\pi^2R k\mathrm{_B} T)$ with $E_\mathrm{C}=e^2/(2C)$ the charging energy.

\subsubsection{Analytical asymptotic expressions versus $V$ at $T=T_\mathrm{node}=T_\mathrm{env}=0$, and versus $T=T_\mathrm{node}=T_\mathrm{env}$ at $V=0$}
\label{DCBtunnelTV0}

We detail here the derivation of analytical expressions at a uniform temperature $T=T_\mathrm{node}=T_\mathrm{env}$ for the asymptotic limits $k_\mathrm{B} T\ll eV\ll \hbar/RC$ (abbreviated in equations as $V\rightarrow 0$, $T=0$), plotted on Fig.~\ref{fig-dcbTunnelV}, and $eV\ll k_\mathrm{B} T \ll \hbar/RC$ (abbreviated as $T\rightarrow 0$, $V=0$), plotted on Fig.~\ref{fig-DCBtunnelT}.

Although the expression given by Eq.~\eqref{PhilEquilibrium} is convenient for performing numerical evaluations in most practical situations, the singularity of the integrand in Eq.~\eqref{PhilEquilibrium} is troublesome when attempting to obtain analytical results. 
To express asymptotic limits in the case of a $RC$ environment, other formulations are required.

(\textit{i}) At $T=T_\mathrm{node}\ll eV/k_\mathrm{B}$, in \cite{Devoret1990}, the conductance at low voltages with respect to the capacitive cutoff ($eV\ll \hbar/RC$) is obtained from a $P(\epsilon)$ formulation such as Eq.~\eqref{eqTunnelP(E)}.
The transmission probability then reads:
\begin{equation}
    \begin{split}
   \frac{\tau(T=0,V\rightarrow 0)}{\tau_\infty}\simeq  \frac{1+2R/R_\mathrm{K}}{\Gamma(2+2R/R_\mathrm{K})}
    \left(\frac{\pi R\,eV}{e^\gamma R_\mathrm{K} E_\mathrm{C}}\right)^{2R/R_\mathrm{K}}.
    \end{split}
    \label{TunnelVtoZero}
\end{equation}

(\textit{ii}) For the linear conductance at $eV\ll k_\mathrm{B}T$, we have developed an equivalent formulation of the tunnel conductance by extending the analysis to the complex plane in $t$, managing the poles and shifting the integral contour. 
We then get:
\begin{equation}
 \frac{\tau(V=0,T)}{\tau_\infty} = \int_0^{+\infty}\!dt\, \frac{\pi k_\mathrm{B} T /\hbar}{\cosh^{2} \frac{\pi t k_\mathrm{B} T}{\hbar}} \times e^{J^\dagger (t,T)}.
\label{Geq2}
\end{equation}
For the simplified $RC$ model of the electromagnetic environment shown in Fig.~\ref{fig-sample}(b), $J^\dagger$ reads:
\begin{equation}
\begin{split}
J^\dagger (t,T)=&\frac{\pi R}{R_\mathrm{K}}\left(\frac{\cos{\frac{\hbar}{2RCk_\mathrm{B} T}}-e^{-t/RC}}{\sin{ \frac{\hbar}{2RCk_\mathrm{B} T}}}-\frac{2tk_\mathrm{B}T}{\hbar}\right)\\
&+\frac{2R}{R_\mathrm{K}}\sum_{n=1}^{\infty}\frac{1-(-1)^n e^{-\omega_nt}}{n(\omega_n^2R^2C^2-1)}.
\end{split}
\label{eqJdagger}
\end{equation}
The sum over Matsubara's frequencies then becomes:
\begin{equation}
    \begin{split}
-\sum_{n=1}^{\infty}\frac{1-(-1)^n e^{-\omega_nt}}{n(\omega_n^2R^2C^2-1)}   =
 \Psi(x)+\frac{1}{2x}+\gamma
+\ln\left(\sqrt{y}+\frac{1}{\sqrt{y}}\right)\\
+\frac{\pi y^x}{2\sin (\pi x )}
-\frac{y}{2(1+x)}~_2 F_1(1,1+x,2+x,-y)\\
- \frac{y}{2(1-x)} ~_2F_1(1,1-x,2-x,-y),
    \end{split}
\label{eqJdaggerMatsubara}
\end{equation}
 recalling for clarity that $\omega_n=2\pi n k_\mathrm{B} T/\hbar$ are Matsubara's frequencies, $\gamma$ is Euler's constant, $\Psi$ is the logarithmic derivative of the Gamma function, $~_2F_1$ is the hypergeometric function, $y=\mathrm{exp}(-2 \pi t k_\mathrm{B} T/\hbar)$, and $x=E_\mathrm{C} R_\mathrm{K}/(2\pi^2R k_\mathrm{B} T)$.

For the asymptotic limit $k_\mathrm{B}T\ll \hbar/RC$, the low temperature behavior is dominated by large values of $t\gg RC$ in Eq.~\eqref{Geq2}. 
We can thus use the long $t$, small $T$ expansion \cite{Ingold1994}: \begin{equation}
J^\dagger(t)\simeq -\frac{2R}{R_\mathrm{K}}\left(\ln\left[2\cosh\left(\frac{\pi t k_\mathrm{B} T}{\hbar}\right)\right]+\ln(2x)+\gamma\right).
\end{equation}
Inserting this expansion in Eq.~\eqref{Geq2}, we find the asymptotic ($eV\ll k_\mathrm{B}T\ll \hbar/RC$) analytical expression of the transmission probability:
\begin{equation}
\begin{split}
 \frac{\tau(V=0,T\rightarrow 0)}{\tau_\infty} \simeq& \frac{\sqrt{\pi}}{2}\frac{\Gamma (1+R/R_\mathrm{K})}{\Gamma (1.5+R/R_\mathrm{K})}\left(\frac{\pi^2 R\,k_\mathrm{B} T}{ e^\gamma R_\mathrm{K} E_\mathrm{C} }\right)^{2R/R_\mathrm{K}}.
\end{split}
\label{LowTeqPowerLaw}
\end{equation} 

Note that previously, in \cite{Iftikhar2016}, we proposed a slightly different empirical expression extracted from \cite{Odintsov1991}, which differs by a factor $(\pi e^ {-2 \gamma})^{R/R_\mathrm{K}}$ from the exact asymptotic expression Eq.~\eqref{LowTeqPowerLaw} (this factor deviates from 1 by less than $1\%$ for $R/R_\mathrm{K}\leq1$).

\subsubsection{Extension to different bath temperatures $T, T_\mathrm{node}, T_\mathrm{env}$}
\label{Appendix3T}

We now focus on the case where the voltage-biased tunnel contact is embedded between two electrodes $L$ and $R$ at different temperatures $T_\mathrm{node}$ and $T$, and with an electromagnetic environment at the temperature $T_\mathrm{env}$.

Following Joyez \textit{et al.} and their notations \cite{Joyez1997}, the relative conductance reduction in the tunnel regime reads:
\begin{equation}
\begin{split}
\frac{\tau}{\tau_\infty}-1=\int{dE}\int{d\epsilon}\, P_\mathrm{Tenv}(\epsilon )f_\mathrm{Tnode}(E-eV)\\
\frac{\partial}{\partial E}\left[f_\mathrm{T}(E+\epsilon )-f_\mathrm{T}(E-\epsilon )\right],
\end{split}
\end{equation}
with $P_\mathrm{Tenv}(\epsilon )$ the probability distribution to exchange the energy $\epsilon$ with the electromagnetic environment (previously introduced in Eq.~\eqref{eqTunnelP(E)}), and $f_{\mathrm{T}x}$ the Fermi function at temperature $T_x$ with $x\in\{L,R\}$.

Equivalently, in the time-domain, the relative conductance reduction reads:
\begin{equation}
\begin{split}
\frac{\tau(V,T_\mathrm{node},T_\mathrm{env},T)}{\tau_\infty}-1=   \int_0^{+\infty}\!dt\, 2\pi t\,\mathrm{Im} \left[e^{J(t,T_\mathrm{env})}\right]\\
\times \frac{ k_\mathrm{B} T/\hbar}{\sinh (\pi t k_\mathrm{B} T/\hbar)} \frac{ k_\mathrm{B} T_\mathrm{node}/\hbar}{\sinh (\pi t k_\mathrm{B} T_\mathrm{node}/\hbar)}\cos \frac{e V t}{\hbar}.
\label{AA3T}
\end{split}
\end{equation} 
In the limit $T=0$, one finds:
\begin{equation}
\begin{split}
\frac{\tau(V,T_\mathrm{node},T_\mathrm{env},T=0)}{\tau_\infty}-1= 2 \int_0^{+\infty}\!dt\, \mathrm{Im} \left[e^{J(t,T_\mathrm{env)}}\right]\\
 \times \frac{k_\mathrm{B} T_\mathrm{node}/\hbar}{\sinh (\pi t k_\mathrm{B} T_\mathrm{node}/\hbar)}\cos \frac{e V t}{\hbar}.
\label{AAoneTnul}
\end{split}
\end{equation} 

A natural approximation for the environment temperature is the average temperature $T_\mathrm{env}=T_\mathrm{node}/2$; if $eV\ll k_\mathrm{B}T_\mathrm{node}$, Eq.~\eqref{AAoneTnul} then simplifies to:
\begin{equation}
\tau(V=0,T_\mathrm{node}=2 T_\mathrm{env},T_\mathrm{env},T=0)/\tau_\infty=e^{J^\dagger (0,T_\mathrm{env})}.
\label{AAoneTnul2}
\end{equation}
Note that Eq.~\eqref{AAoneTnul2} is equivalent in the energy domain to:  
\begin{equation}
\begin{split}
   &\tau(V=0,T_\mathrm{node}=2 T_\mathrm{env},T_\mathrm{env},T=0)/\tau_\infty=\\
   &\int_{-\infty}^\infty d\epsilon\frac{2P_{\mathrm{Tenv}}(\epsilon)}{1+e^{\epsilon/2k_\mathrm{B}T_\mathrm{env}}}=
  \int_{-\infty}^\infty d\epsilon P_{\mathrm{Tenv}}(\epsilon)e^{-\epsilon/2k_\mathrm{B}T_\mathrm{env}}.
    \end{split}
\end{equation}

When $T_\mathrm{env}=T_\mathrm{node}/2\ll \hbar/k_\mathrm{B}RC$, we have $J^\dagger(0,T_\mathrm{env})\simeq -\frac{2R}{R_\mathrm{K}}\left(\ln(2x)+\gamma \right)$.
The low temperature asymptotic behavior of Eq.~\eqref{AAoneTnul2} then reads :
\begin{equation}
\begin{split}
\tau(V=0,T_\mathrm{node}\ll \hbar/k_\mathrm{B}RC,T_\mathrm{env}=T_\mathrm{node}/2,T=0)/\tau_\infty
\simeq\\
\left(\frac{\pi^2 R\,k_\mathrm{B} T_\mathrm{node} }{2\,e^\gamma R_\mathrm{K} E_\mathrm{C}}\right)^{2R/R_\mathrm{K}}.
\label{LowToneNulPowerLaw}
\end{split}
\end{equation}

Equation~\ref{LowToneNulPowerLaw} with one null temperature and Eq.~\eqref{LowTeqPowerLaw} with uniform temperatures are both temperature power laws with the same exponent $2R/R_\mathrm{K}$.
They correspond to, respectively, the red and black dashed lines plotted on Fig.~\ref{fig-DCBtunnelT}.
The constant ratio between these limits allows us to calculate the temperature rescaling factor $\alpha$ discussed in the main paper:
\begin{equation}
    \begin{split}
       \frac{\tau(V=0,T_\mathrm{node}\ll \hbar/k_\mathrm{B}RC, T_\mathrm{env}=T_\mathrm{node}/2,T=0)}{\tau(V=0,T_\mathrm{node}=T_\mathrm{env}=T\ll\hbar/k_\mathrm{B}RC)}=\\
       \frac{2}{\sqrt{\pi}}\frac{\Gamma(1.5+R/R_\mathrm{K})}{\Gamma(1+R/R_\mathrm{K})}2^{-2R/R_\mathrm{K}}
       =\alpha^{2R/R_\mathrm{K}}.
    \end{split}
\end{equation}
Consequently, the temperature reduction factor $\alpha$ reads, in the tunnel regime:
\begin{equation}
  \alpha = \frac{1}{2}\left[\frac{2}{\sqrt{\pi}}\frac{\Gamma(1.5+R/R_\mathrm{K})}{\Gamma(1+R/R_\mathrm{K})}\right]^{R_\mathrm{K}/2R}.
    \label{eq-rescaledTI}
\end{equation}
For the implemented series resistances $R=R_\mathrm{K}/2,$ $R_\mathrm{K}/3$ and $R_\mathrm{K}/4$, we obtain $\alpha\simeq0.637,$ $0.648$ and $0.655$, respectively. 

\subsection{Dynamical Coulomb blockade theory in the near-ballistic regime}

\subsubsection{Quantitative predictions at a uniform temperature $T=T_\mathrm{node}=T_\mathrm{env}$}
\label{CMballistic}

The power law exponent in the vicinity of the ballistic regime is known from the duality predicted between strong back-scattering (tunnel) and weak back-scattering (near ballistic) regimes across an impurity in a Luttinger liquid of interaction parameter $K$ and $1/K$, respectively \cite{Kane1992b,Fendley1998,Lesage1999}.
The corresponding prefactor is nonetheless not universal and depends on the microscopic details, such as the high frequency capacitive cutoff. 
It has been inferred in the particular case $R=R_\mathrm{K}$ ($K=1/2$) in \cite{Anthore2018}, adapting \cite{Furusaki1995b}.
We extend here such a quantitative prediction to $R=R_\mathrm{K}/N$ with $N\in\mathbb{N}$.
Remarkably, we find that the duality also exactly applies to the prefactor.

Following Ref.~\cite{Furusaki1995b}, we assume an energy-independent back-scattering at the contact in the absence of DCB, and describe the $N$ fully ballistic channels and the weakly reflected one with bosonic variables $\phi_j (x)$ ($j=1,\ldots,N+1$). 
The nearly ballistic (weakly reflected) edge channel is denoted by $j=1$.
Each channel has a fictitious right(left)-moving part, corresponding to the edge path before (after) entering the charged island in the region $x>0$. 
The island electric charge is thus
\begin{equation}
    \hat{Q} = - \frac{e}{\pi} \sum_{j=1}^{N+1} \int_0^{+\infty} d x \partial_x \phi_j (x) = \frac{e}{\pi} \sum_j \phi_j,
\end{equation}
where we set the notation $\phi_j \equiv \phi_j (0)$. 
It is in fact easier to work with the (properly normalized) total charge field $\tilde{\phi}_1 = \frac{1}{\sqrt{N+1}} \sum_j \phi_j$. Employing current conservation, the Kubo formula can be written as
\begin{equation}
    G = G_{\rm max} \frac{2 \omega_n}{\pi} \, \langle \tilde{\phi}_2 (\omega_n) \tilde{\phi}_2 (-\omega_n) \rangle_{i \omega_n \to 0^ +},
\end{equation}
where the imaginary (Matsubara) frequency $\omega_n = 2 \pi n k_\mathrm{B}T/\hbar$ is analytically continued to the real axis and then sent to zero.
Here we have introduced a second linear combination of the original fields $\tilde{\phi}_2 = \sqrt{N/(N+1)} \left( \phi_1 - \frac{1}{N} \sum_{j \ne 1} \phi_j \right)$, with coefficients orthogonal to those of $\tilde{\phi_1}$. The advantage of performing this orthogonal change of variables is that the Hamiltonian then couples only the two fields $\tilde{\phi}_1$ and $\tilde{\phi}_2$. We can factor out the other linear combinations $\tilde{\phi}_j$ ($j\ge 2$) for the evaluation of the Kubo formula.

With this formulation, the Euclidean action that governs the dynamics of the two relevant bosonic fields is $S = \sum_{j=1,2} \sum_{n=0}^{+\infty}  \tilde{\phi}_j (i \omega_n) K_j \tilde{\phi}_j (-i \omega_n) + S_{\rm BS}$ with the inverse Green's functions $\pi K_j = |\omega_n| + \delta_{j,1} (N+1) E_\mathrm{C}/\pi $ and the back-scattering term
\begin{equation}
    S_{BS} = \frac{D \sqrt{1 - \tau_\infty}}{\pi} \! \! \int_0^{ \hbar/k_\mathrm{B} T} \! d \tau \cos \left( \frac{2 \tilde{\phi}_1(\tau) +2 \sqrt{N}\tilde{\phi}_2 (\tau) }{\sqrt{N+1}}    \right) ,
\end{equation}
with $D$ the edge electrons' energy bandwidth that is necessarily introduced in bosonization.
$D$ acts as a high-energy regularization which cancels out when evaluating the conductance.
 
Equipped with this action, we follow Appendix~A.1 from Ref.~\cite{Furusaki1995b} and compute the conductance to leading non-vanishing order in the back-scattering amplitude $\sqrt{1 - \tau_\infty} \ll 1$. We find the analytical result
\begin{equation}
    \begin{split}
    G_\mathrm{max}-G=&
   \frac{R_\mathrm{K}\sqrt{\pi}\left( 1-\tau_\mathrm{\infty}\right)}{2\left(R+R_\mathrm{K}\right)^2}
 \frac{\Gamma\left[R_\mathrm{K}/\left(R+R_\mathrm{K}\right)\right]}{\Gamma\left[1/2+R_\mathrm{K}/\left(R+R_\mathrm{K}\right)\right]}
\\
   &\times \left[\frac{e^\gamma  E_\mathrm{C} (1+R_\mathrm{K}/R)}{\pi^2 k_\mathrm{B}T}\right]^{\frac{2 R}{R+R_\mathrm{K}}},
    \end{split}
    \label{eqGmaxMinusGNearBallistic}
\end{equation}
where we recall that $G_{\rm max} = (R+R_K)^{-1}$. 
We obtain the desired temperature scaling  $T^{2 K -2}$ also predicted from the duality tunnel-near ballistic with, in addition, an exact prediction for the prefactor in terms of the transmission $\tau_\infty$, the charging energy $E_\mathrm{C}=e^2/2 C$ and the ratio of resistances $R_\mathrm{K}/R$ corresponding to the number of ballistic channels connecting the island.

Remarkably, although this was not expected to our knowledge, we find that the duality between tunnel and near ballistic regimes also applies for the exact value of the multiplicative factor, despite the dependence of this prefactor on the capacitive cutoff.
More precisely, we compare the expressions of $\tau/\tau_\infty$ in the tunnel regime (obtained from Eq.~\eqref{LowTeqPowerLaw}), with $(G_\mathrm{max}-G)/(G_\mathrm{max}-G_\infty)$ in the near ballistic regime  (obtained from Eq.~\eqref{eqGmaxMinusGNearBallistic}), where $G_\infty\equiv(R_\mathrm{K}/\tau_\infty+R)^{-1}$ is the device conductance in the absence of DCB renormalization.
It turns out that these two expressions map \textit{exactly} onto one another provided $K=R_\mathrm{K}/(R+R_\mathrm{K})$ is replaced by $1/K=R/R_\mathrm{K}+1$.
This remarkable robustness of the duality also suggests that Eq.~\eqref{eqGmaxMinusGNearBallistic}, which was obtained for $R=R_\mathrm{K}/N$, may apply for arbitrary values of $R$ (a theoretical treatment of arbitrary $R$ is in preparation \cite{Safi_preparation}).

\subsubsection{Comparison quantitative predictions-experiments at a uniform temperature $T=T_\mathrm{node}=T_\mathrm{env}$}
\label{ballisticCMvsData}
\begin{figure}
    \centering
    \includegraphics[width=\columnwidth]{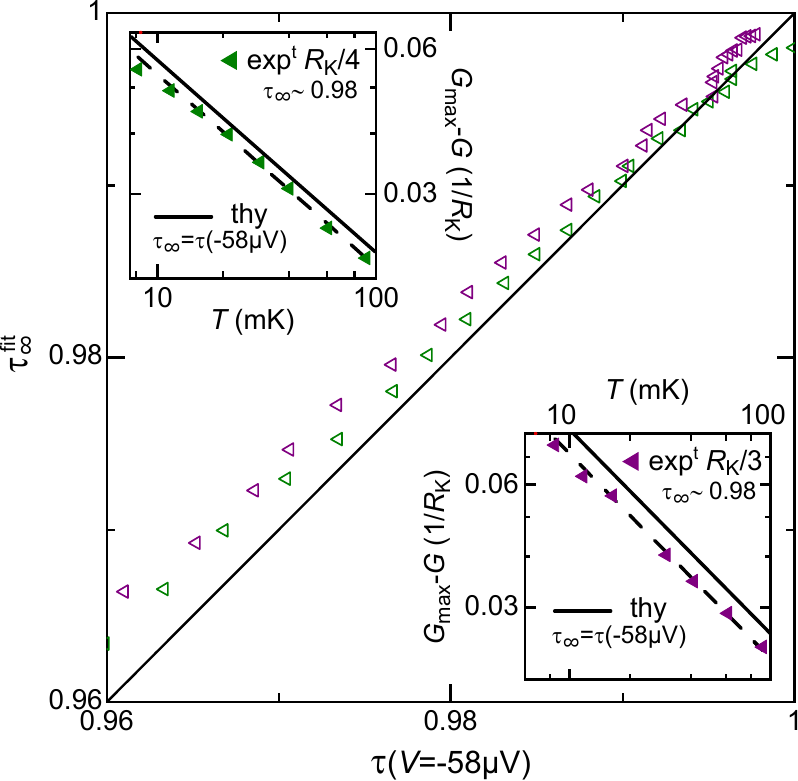}
    \caption{Insets: symbols show illustrative measurements at $\tau_\infty\sim0.98$ of $G-G_\mathrm{max}$ for $R=R_\mathrm{K}/N$ (top and bottom panel: $N=4$ and $3$, resp.) versus equilibrium temperature $T$ at $V=0$, with $G_\mathrm{max}=\frac{1}{R_\mathrm{K}}\frac{N}{N+1}$ (data also shown in Fig.~\ref{fig-tauClose1T});
    Continuous lines are quantitative theoretical predictions of Eq.~\eqref{eqGmaxMinusGNearBallistic}, using $\tau_\infty=\tau(V=-58\,\mu$V) without any fit parameter;
    Dashed lines are fits using $\tau_\infty$ as a free parameter adjusted by matching the data point at $T=90$\,mK. 
    Main panel: symbols represent the fitted values of $\tau_\infty$ versus the corresponding large bias voltage measurements $\tau(V=-58\,\mu$V).}
    \label{fig-SI-Close1PowerLaw}
\end{figure}

In Fig.~\ref{fig-SI-Close1PowerLaw}, we confront the new predictions of Eq.~\eqref{eqGmaxMinusGNearBallistic} with the experimental conductance measured at equilibrium in the near ballistic regime, for $R=R_\mathrm{K}/3$ and $R_\mathrm{K}/4$.

The insets display a direct comparison of the predicted (continuous lines) and measured (symbols) conductance at a representative channel tuning of $\tau_\infty\sim0.98$.
The quantitative predictions of Eq.~\eqref{eqGmaxMinusGNearBallistic} are calculated without any fit parameter, assuming $\tau_\mathrm{\infty}\simeq\tau(V=-58\,\mu\mathrm{V})$ on the basis that for such large dc bias voltage only a relatively small renormalization due to DCB is expected ($V$ being of the order of the capacitive cutoff $N E_\mathrm{C}/ \pi e=h/2\pi eRC$).
We observe here a relatively small quantitative discrepancy of $\sim 7\%$ and $\sim16\%$ for $R=R_\mathrm{K}/4$ and $R_\mathrm{K}/3$, respectively.
This small discrepancy could result from the experimental uncertainty on $\tau_\infty$, due to a residual DCB renormalization as well as a non-negligible energy dependence of $\tau_\infty$ at large bias voltages.

In the main panel, we perform a quantitative data/theory comparison over a broad span of $\tau_\infty\in[0.96,1]$.
For this purpose, the fitted value $\tau_\mathrm{\infty}^\mathrm{fit}$ is obtained by matching the prediction of Eq.~\eqref{eqGmaxMinusGNearBallistic} with the conductance measured at $T\simeq90$\,mK. 
The resulting $\tau_\mathrm{\infty}^\mathrm{fit}$ is plotted as symbols versus the measured transmission probability at high bias voltage $\tau(V=-58\,\mu\mathrm{V})$.
In the ideal case where $\tau_\mathrm{\infty}=\tau(V=-58\,\mu\mathrm{V})$, we would expect the $\tau_\mathrm{\infty}^\mathrm{fit}$ points to fall on the continuous straight line corresponding to  $\tau_\mathrm{\infty}^\mathrm{fit}=\tau(V=-58\,\mu\mathrm{V})$.
We observe that the data points are relatively close to this line, and that the distance reduces as $\tau_\infty$ approaches one. 
This comparison establishes the quantitative predictions of Eq.~\eqref{eqGmaxMinusGNearBallistic} at a good relative accuracy, which we believe is here limited by experimental discrepancies between $\tau_\infty$ and $\tau(V=-58\,\mu\mathrm{V})$.

\subsubsection{Near ballistic theory with different bath temperatures $T, T_\mathrm{node}, T_\mathrm{env}$}
\label{ISballistic3T}

The QPC is here coupled to two electrodes at temperatures $T_\mathrm{node}$ and $T$. 
In that case one cannot use the Euclidian action employed at equilibrium, and an adapted Keldysh approach is required.
Here, we restrict ourselves to a simple resistance $R=R_\mathrm{K}/N$ in series with the QPC. 
A full analysis including exactly the parallel capacitance $C$ will be performed separately \cite{Safi_preparation}.

The QPC is modeled by a weak local back-scattering term at $x=0$:
\begin{equation}
    H_\mathrm{BS}=\sqrt{1-\tau_{\infty}}\cos(2\phi_1(0))/2\pi t_0,
\end{equation}
with $1-\tau_{\infty}\ll 1$, $t_0$ a short time cutoff of the order of $\hbar/E_\mathrm{C}$, and $\phi_1$ the bosonic field introduced in Appendix~\ref{CMballistic}. 
We treat the remaining $N$ channels as a linear resistance $R=R_\mathrm{K}/N$ (note that the present approach applies to arbitrary values of $R$). 
The coupling term between the QPC and this environment reads: $e\phi_1(0)(V-\hat{u})/\pi$, with $V$ the voltage applied to the all device (QPC and series resistance), $e\phi_1(0)/{\pi}$ the total charge transferred through the QPC, and $\hat{u}$ the voltage operator across the resistance, whose fluctuations are given by $\partial_{t}^2J(t)$ (see Eq.~\eqref{eqJ(t)}) and are determined by $R$ and $T_\mathrm{env}$.

First, we need to distinguish the right and left going electron fields $\Psi_{R,1}, \Psi_{L,1}$.
$\Psi_{R,1}$ moves away from the island, at a temperature $T_\mathrm{node}$.
$\Psi_{L,1}$ moves toward it, corresponding to electrons injected from the right electrode at temperature $T$ (see article Fig.~\ref{fig-sample}(b)).
Second, we adopt a similar strategy to Ref.~\cite{Safi2004} by integrating out the environment, ending up with an effective Keldysh action for the bosonic field $\phi_1(0)$ \cite{Safi_preparation}.

The mapping of this DCB problem to a one-dimensional Luttinger liquid with an impurity breaks down (also when including $C$, which corresponds to finite-range interactions as discussed in the supplemental material of \cite{Jezouin2013}).
Yet, it is still convenient to use the parameter $K=(1+R/R_\mathrm{K})^{-1}$, which  determines the non-diagonal element of the Keldysh matrix Green's function for $\phi_1(0)$:
\begin{eqnarray}\label{eqC_neq}
C^\mathrm{neq}(t;T_\mathrm{node},T_\mathrm{env},T)&=&\frac{K}2\left[C^\mathrm{eq}(t;T_\mathrm{node})+C^\mathrm{eq}(t;T)\right]\nonumber\\&&+(1-K)C^\mathrm{eq}(t;T_\mathrm{env}).
\end{eqnarray}
Here we use the Green's function obtained at a uniform temperature $T$ in a Luttinger liquid with parameter $K$ and at the same cutoff $t_0$:  $C^\mathrm{eq}(t;T)= -(K/2)\ln{ \left[\hbar \sinh\left( \pi k_\mathrm{B}T(-t+it_0)/\hbar\right)/\pi k_\mathrm{B}Tt_0)\right]}$.  

To lowest order with respect to the back-scattering amplitude $\sqrt{1-\tau_{\infty}}$, the Green's function $C^\mathrm{neq}(t;T_\mathrm{node},T_\mathrm{env},T)$ determines fully the current as a function of the voltage $V$ and temperatures $T_\mathrm{node},T_\mathrm{env},T$.
Here we restrict ourselves to the experimentally measured linear conductance at zero dc voltage, using an extension of Kubo's formula \cite{ines_philippe,ines_PRB_2019}. 
We find:
\begin{eqnarray}\label{X_explicit}
G_\mathrm{max}-G(T_\mathrm{node},T_\mathrm{env},T)=&\nonumber\\-iK^2\frac{1-\tau_{\infty}}{\pi t_0^2 R_\mathrm{K}}\int_{-\infty}^{\infty}\!\!dt\; t\; &e^{4C^\mathrm{neq}\!(t;T_\mathrm{node},T_\mathrm{env},T)}.
\end{eqnarray}
We can determine the effective time cutoff $t_0$ by comparison to the prediction for a uniform temperature $T$ given in Eq.~\eqref{eqGmaxMinusGNearBallistic}: $t_0=\hbar\pi Ke^{-\gamma}/E_\mathrm{C}$, with $\gamma$ the Euler's constant.
Note that this prefactor can also be recovered from a complete analysis including $C$, as will be detailed elsewhere \cite{Safi_preparation}.

Injecting Eq.~\eqref{eqC_neq} into Eq.~\eqref{X_explicit} and restricting ourselves to the experimental hypothesis $T_\mathrm{env}=(T+T_{\mathrm{node}})/2$, we finally obtain:
\begin{equation}
\begin{split}
G_\mathrm{max}-G(T_\mathrm{node},\frac{T_\mathrm{node}+T}{2},T)= \frac {2K^2(1-\tau_{\infty})}{\pi R_\mathrm{K}} \sin \pi K \\
   \times \left[\frac{2e^{\gamma} E_\mathrm{C}}{\pi^2 K k_\mathrm{B} (T_\mathrm{node}+T)}\right]^{2(1-K)}
   \left[\frac{8\;T_\mathrm{node}T}{(T_\mathrm{node}+T)^2}\right]^{K^2} \\
  \int_0^{\infty}\! dt\frac{t}{\left(\sinh t\right)^{2K(1-K)}}\left[\cosh 2t-\!\cosh \left(2t\frac{T_\mathrm{node}-T}{T_\mathrm{node}+T}\right)\right]^{-K^2}.
\end{split}
\label{eqDTweakbs}
\end{equation}
We recall that $K=(1+R/R_\mathrm{K})^{-1}$ and $G_{\rm max} = (R+R_\mathrm{K})^{-1}=K/R_\mathrm{K}$.
The integral in Eq.~\eqref{eqDTweakbs} can be readily evaluated numerically.
This allows one to compute the conductance at arbitrary values of $T$, $T_\mathrm{node}$, as long as both remain small with respect to the high energy cutoff ($T,T_\mathrm{node}\ll E_\mathrm{C}/k_\mathrm{B}$) and that the back-scattering remains weak ($1-G/G_\mathrm{max}\ll1$).
The continuous lines in Fig.~\ref{fig-tauClose1T} are the predictions of Eq.~\eqref{eqDTweakbs}.

\subsection{Experimental electronic temperatures}
\label{App-NoiseThermometry}

Having a good knowledge of the different electronic temperatures (base $T$ and node  $T_\mathrm{node}$) is crucial for the present experiments.
In this section, we first summarize how these temperatures are separately measured.
Then we detail how $T_\mathrm{node}$ can be calculated based on the heat Coulomb blockade theory previously established, and compare with our measurements.
Finally, we discuss the possible choices for the temperature $T_\mathrm{env}$ of the electromagnetic environment composed of the series $RC$ circuit.

\subsubsection{Measurement of the electrons' temperature $T$ in the large electrodes}
Following \cite{Iftikhar2016}, we have obtained $T$ from shot noise measurements in a device configuration where the metallic island is bypassed (thanks to lateral gates visible in Fig.~\ref{fig-sample}(a), which are operated as short-circuit switches).
For temperatures $T\geq40$\,mK, we used the mixing chamber temperature measured by a RuO$_2$ thermometer, which was previously shown to match very closely the electrons' temperature on the same setup \cite{Iftikhar2016}.

\subsubsection{Measurement of the electrons' temperature in the metallic island $T_\mathrm{node}$}

Following \cite{Sivre2019}, the temperature increase of the electrons in the central node is inferred from two independent noise measurements, performed on the electrodes 1 and 3 of Fig.~\ref{fig-sample}(a). 
Here, one noise measurement is realized behind the partially transmitted channel (on electrode numbered $3$, schematically connected to an amplifier and resonator) and the other one behind the $N_1$ ballistic channels (on electrode numbered $1$, also schematically connected to an amplifier).
Specifically, we measure the difference with respect to equilibrium in the auto-correlation signals $\Delta S_{11}$ and $\Delta S_{33}$, and in the cross-correlation signal $\Delta S_{13}$.
From current conservation and the negligible charge accumulation in the device at the MHz measurement frequencies, we find following \cite{Sivre2018,Sivre2019}  that the thermal noise increase $\Delta S^\mathrm{th}\equiv 2 k_\mathrm{B}(T_\mathrm{node}-T)/R_\mathrm{K}$ is given by the excess (increase in) noise signals:
\begin{equation}
    \Delta S^\mathrm{th}=\Delta S_{11}\frac{N_1+N_2}{N_1N_2}-\Delta S_{33}\frac{N_1}{N_2\left(N_1+N_2\right)}
\end{equation}
or alternatively  
\begin{equation}
    \Delta S^\mathrm{th}=\Delta S_{11}\frac{N_1+N_2}{N_1N_2}+\frac{\Delta S_{13}}{N_2}.
\end{equation}
Note that both expressions allow one to extract $T_\mathrm{node}-T$ independently. 
We have checked that they were equivalent.
None of these expressions depend on $\tau$.

In practice, each data point is averaged over about $10$\,min to get a temperature resolution of $\sim 0.1$\,mK.
At this resolution, we are also sensitive at our lowest temperature to a small heating of the central node by the spurious low-frequency noise induced by vibrations. 
This noise, which depends on the device configuration, is separately determined to be $\delta V_\mathrm{noise}\sim 0.4\,\mathrm{\mu V}$.
It results in a small heating of the central node of $\lesssim 0.2$\,mK.

\begin{figure}
    \centering
    \includegraphics[width=\columnwidth]{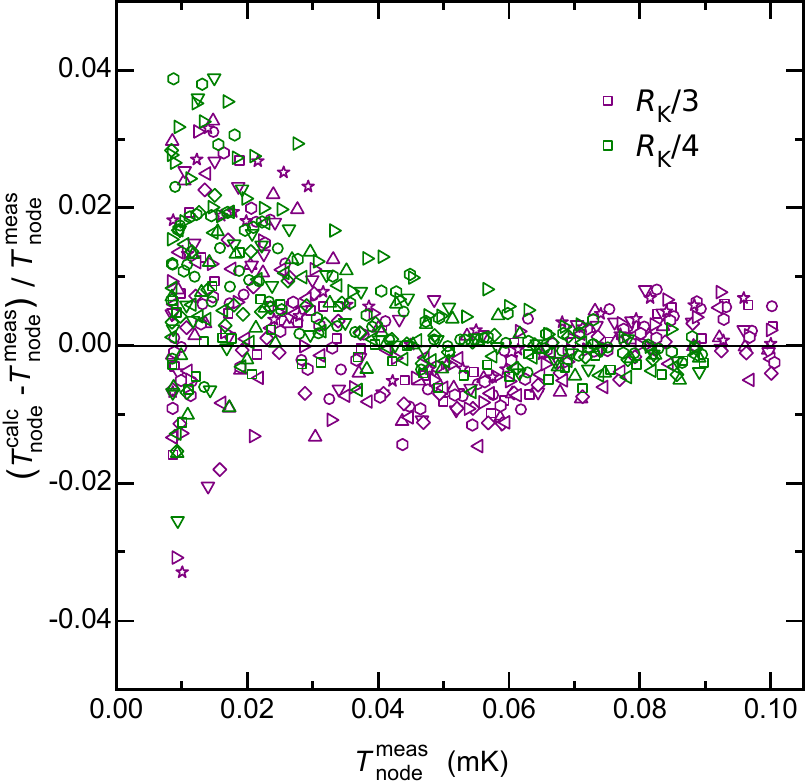}
    \caption{Symbols: relative difference between the calculated node temperature $T_\mathrm{node}^\mathrm{calc}$ and the measured one $T_\mathrm{node}^\mathrm{meas}$, shown at base temperature $T\sim 8$\,mK for both $R=R_\mathrm{K}/3$ and $R=R_\mathrm{K}/4$, over the full range of explored $\tau$ values.}
    \label{fig-siTcalcVersusTmeas}
\end{figure}

\subsubsection{Calculation of the electrons' temperature in the metallic island $T_\mathrm{node}$}

We also relied on our knowledge of heat flow in the device \cite{Sivre2019} to calculate $T_\mathrm{node}$.
These calculated values were used in the tunnel regime and also for the out-of-equilibrium measurements performed at temperatures $T$ larger than our base temperature $\sim8$\,mK.

The node temperature is determined by balancing the injected Joule power in the metallic node ($P_\mathrm{J}$) with the outgoing heat currents, from electrons to phonons ($J_\mathrm{ph}^Q$) and through the connected electronic channels ($J_\mathrm{el}^Q$).

The electron-phonon heat flow is determined when the device only hosts ballistic channels.
We find:
\begin{equation}
    J_\mathrm{ph}^Q\simeq 1.8 \times 10^{-8}\left(T_\mathrm{node}^{5.5}-T^{5.5}\right)\,\mathrm{W}.
\end{equation}

The flow of heat across the electronic channels reads \cite{Sivre2019}:
\begin{equation}
\begin{split}
J_\mathrm{el}^Q=(& N  + \tau) \frac{\pi^2k_\mathrm{B}^2}{6h}\left(T_\mathrm{node}^{2}-T^{2}\right)\\
+&(N+\tau)\frac{(N+\tau^2) E_\mathrm{C}^2}{\pi^2 h}\\
&\times\left[I\left(\frac{(N+\tau)E_\mathrm{C}}{\pi k_\mathrm{B} T}\right)-I\left(\frac{(N+\tau)E_\mathrm{C}}{\pi k_\mathrm{B} T_\mathrm{node}}\right)\right],
\end{split}
\end{equation}
with $I(x)=\frac{1}{2}\left[\ln{\left(\frac{x}{2\pi}\right)}-\frac{\pi}{x}-\psi\left(\frac{x}{2\pi}\right)\right]$.

Knowing the injected power $P_\mathrm{J}=\frac{N_1 V_1^2 + N_2 V_2^2}{2R_\mathrm{K}}$, $N=N_1+N_2$, the charging energy $E_\mathrm{C}=k_\mathrm{B}\times370$\,mK and the temperature $T$, we can solve the heat balance equation for each measured point $\tau$ and thereby find the only unknown parameter $T_\mathrm{node}$. 

The relative accuracy of the heat Coulomb blockade theory on the present device is tested at base temperature $T\sim8\,$mK in Fig.~\ref{fig-siTcalcVersusTmeas}, where we plot as symbols the relative difference between calculated and measured node temperatures. 
The agreement is better than $4\%$ over the full $\tau$ and $T_\mathrm{node}$ ranges.

\begin{figure}
    \centering
    \includegraphics[width=\columnwidth]{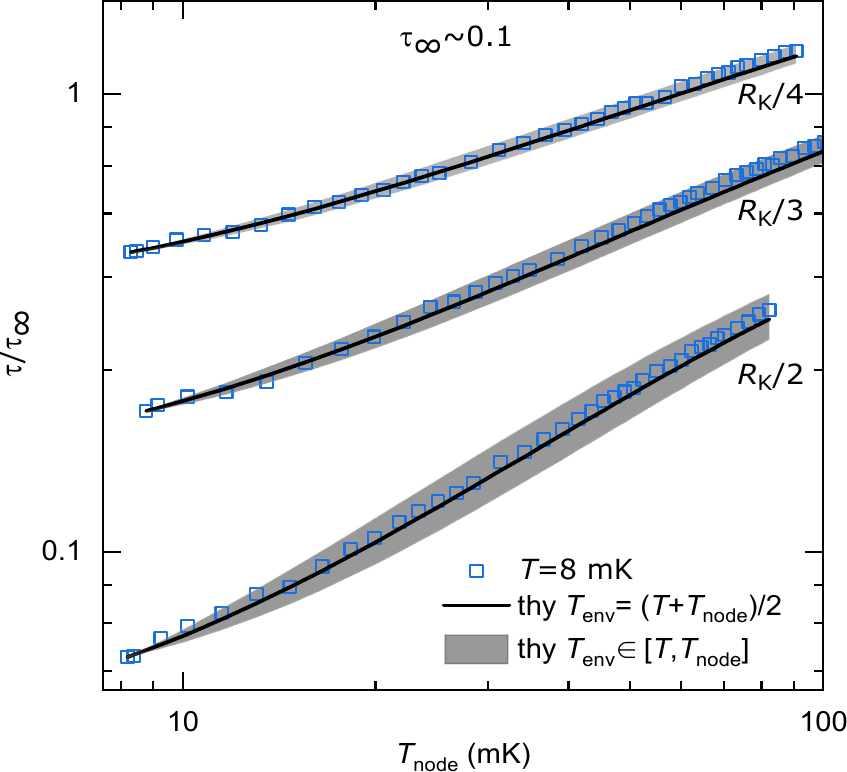}
    \caption{Open symbols: renormalized transmission probability $\tau/\tau_\infty$ of the generic channel in series with a resistance $R_\mathrm{K}/2$, $R_\mathrm{K}/3$ and $R_\mathrm{K}/4$ versus the node temperature at $V=0$ at the base temperatures $T\simeq8$\,mK in a log-log scale.
    Black lines: predictions of the full tunnel DCB theory for different temperatures (see Appendix~\ref{Appendix3T}) calculated using $T=8$\,mK, $T_\mathrm{node}$, and $T_\mathrm{env}=(T+T_\mathrm{node})/2$.
    $\tau_\infty$ is the only adjustable parameter per value of $R/R_\mathrm{K}$ in the data-theory comparison. 
    The gray areas correspond to the predicted range of conductance for $T_\mathrm{env}\in [T,T_\mathrm{node}]$, using $T=8$\,mK and $T_\mathrm{node}$.}
    \label{fig-SI-DCBTunnel-Tenv}
\end{figure}

\subsubsection{Temperature of the electromagnetic environment $T_\mathrm{env}$}
\label{AppendixTenv}

The environment temperature appears as a separate parameter $T_\mathrm{env}$ in the tunnel DCB theory as well as in the novel theory developed in the near ballistic regime (Appendix~\ref{ISballistic3T}).
In the main paper, we use the average value $T_\mathrm{env}=(T_\mathrm{node}+T)/2$.
Here, we determine the range of $T_\mathrm{env}$ over which the tunnel DCB theory is compatible with the data.

For this purpose, we show in Fig.~\ref{fig-SI-DCBTunnel-Tenv} the same tunnel data points as in Fig.~\ref{fig-DCBtunnelT}, and the black continuous lines also correspond to the tunnel DCB theory predictions with $T_\mathrm{env}=(T_\mathrm{node}+T)/2$.
In addition, the gray areas enclose the tunnel DCB predictions for the full interval $T_\mathrm{env}\in [T,T_\mathrm{node}]$.
In practice the data points are close or slightly above the prediction for $T_\mathrm{env}=(T_\mathrm{node}+T)/2$, suggesting that $T_\mathrm{env}\gtrsim(T_\mathrm{node}+T)/2$.
However, the interval $T_\mathrm{env}\in [\sim(T_\mathrm{node}+T)/2,T_\mathrm{node}]$ remains within our experimental uncertainty (approximately the size of the points).

\subsection{Complementary data}

\subsubsection{DCB under a bias voltage in the tunnel regime at $R_\mathrm{K}/2$ and $R_\mathrm{K}/4$}

\begin{figure}[htb!]
    \centering
    \includegraphics[width=\columnwidth]{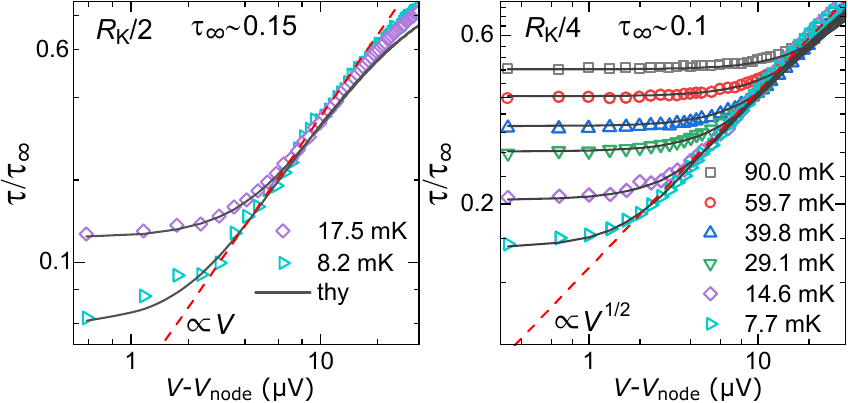}
    \caption{Tunnel DCB theory-data comparison under a bias voltage at $R_\mathrm{K}/2$ and $R_\mathrm{K}/4$.
    Symbols: measured renormalized transmission probability across the generic channel in series with a resistance $R_\mathrm{K}/2$ ($R_\mathrm{K}/4$) plotted versus the channel bias voltage $V-V_\mathrm{node}$ at different temperatures $T$ in a log-log scale.
    Black lines: full tunnel DCB theory calculated with the parameters $C=3.1$\,fF ($C=2.5$\,fF) and the measured temperature $T$.
    Red dashed line: asymptotic power law predictions at zero temperature with no fit parameter.}
    \label{fig-SI-DCBTunnel-V}
\end{figure}

To complement Fig.~\ref{fig-dcbTunnelV} focusing on $R=R_\mathrm{K}/3$, we plot in Fig.~\ref{fig-SI-DCBTunnel-V} the measured conductances (symbols) and DCB predictions (lines) in the tunnel regime for the series resistance $R=R_\mathrm{K}/2$ and $R_\mathrm{K}/4$ at different temperatures $T$.

\subsubsection{Full Tomonaga-Luttinger conductance renormalization curve at equilibrium}

\begin{figure}[htb!]
    \centering
    \includegraphics[width=0.9\columnwidth]{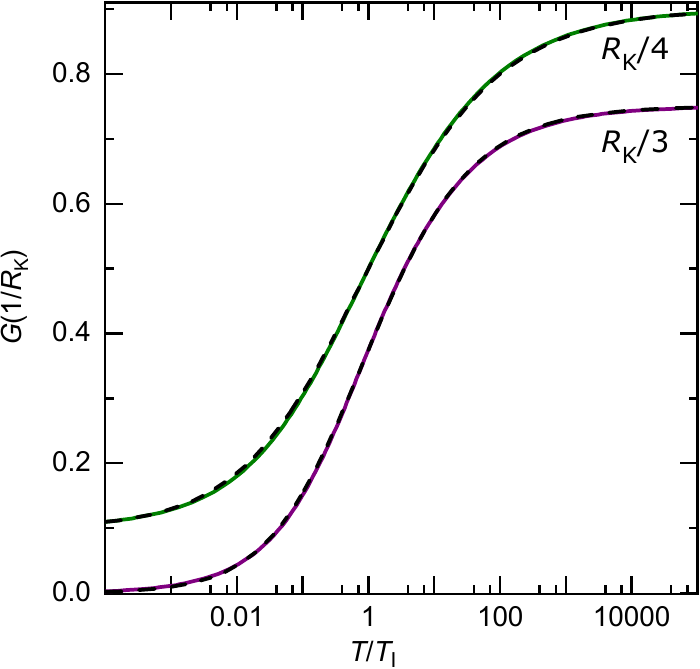}
    \caption{Universal renormalization flow of the conductance at equilibrium. 
    Colored continuous lines represent the experimental curves (green for $R_\mathrm{K}/4$ is shifted vertically by $0.1$, purple shows $R_\mathrm{K}/3$) obtained by averaging the ensemble of data measured from $T=8$\,mK to $T=90$\,mK for each $\tau_\infty$ configuration (see \cite{Anthore2018} for the detailed procedure).
    The exact theoretical predictions $G^\mathrm{eq}_{R/R_\mathrm{K}}(T/T_\mathrm{I})$ derived in \cite{Boulat2020} are shown as black dashed lines.
    The full renormalization curve has not been measured for $R=R_\mathrm{K}/2$.}
    \label{fig-SI-HomodGsdLnT}
\end{figure}

As in \cite{Anthore2018,Anthore2020}, we show on Fig.~\ref{fig-SI-HomodGsdLnT} the pertinence of the mapping to a TLL by comparing the measured conductance $G(V=0,T)$ versus $T/T_I$ at equilibrium (colored continuous lines) to the predictions $G^\mathrm{eq}_{R/R_\mathrm{K}}(T/T_\mathrm{I})$ from \cite{Boulat2020} (black dashed lines). 

\begin{figure*}[htb!]
    \centering
    \includegraphics[width=\textwidth]{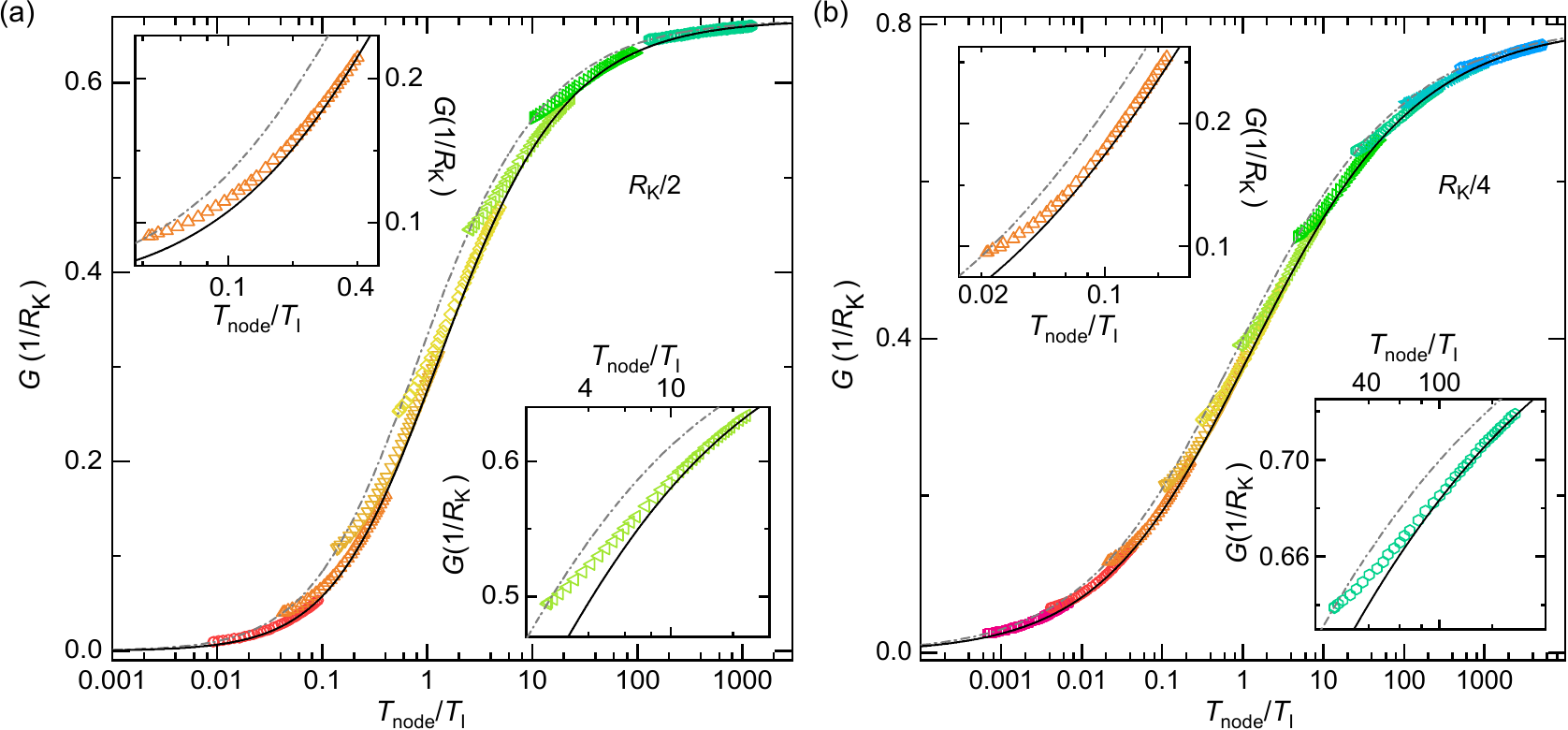}
    \caption{Conductance of a generic channel in series with $R=R_\mathrm{K}/2$ (panel (a)) and $R_\mathrm{K}/4$ (panel (b)) under a temperature bias ($T_\mathrm{node}\geq T$) at base temperature $T\simeq8$\,mK.
    Open symbols: measured sample conductance $G$ for different channel settings of $\tau_\infty$, each shown using a different color and symbol shape.
    For a given channel setting, a unique value of the renormalization temperature $T_\mathrm{I}$ is determined by matching the first data point at equilibrium ($T_\mathrm{node}\simeq T$) with the predicted universal conductance curve at equilibrium $G^\mathrm{eq}_{R/R_\mathrm{K}}(T_\mathrm{node}/T_\mathrm{I})$ (gray dash-dotted line). 
    Black lines: universal conductance curve at equilibrium with the same effective reduction in temperature expected at large $T_\mathrm{node}/T$ from the tunnel DCB theory, namely $G^\mathrm{eq}_{R/R_\mathrm{K}}(\alpha T_\mathrm{node}/T_\mathrm{I})$ with $\alpha$ given in Eq.~\eqref{eq-rescaledTI}.}
    \label{fig-SI-FullRKs2RKs4}
\end{figure*}

\subsubsection{DCB of a generic channel under a temperature bias with $R_\mathrm{K}/2$ and $R_\mathrm{K}/4$}

The figure \ref{fig-SI-FullRKs2RKs4} complements the data at arbitrary channel tuning shown in main manuscript Fig.~\ref{fig-GeneralTau} for $R=R_\mathrm{K}/3$, with here the conductance measured at $R_\mathrm{K}/2$ (panel (a)) and $R_\mathrm{K}/4$ (panel (b)).

\end{document}